\documentclass[a4paper,pra,twocolumn,superscriptaddress]{revtex4-1}

\usepackage{amsmath,amsfonts,amssymb,bm}
\usepackage{nicefrac}

\newcommand{\ket}[1]{\vert#1\rangle}
\newcommand{\bra}[1]{\langle#1\vert}
\newcommand{\Id}{{\mathbb I}}

\usepackage{graphicx}
\newcommand{\figg}[2]{\ensuremath{\vcenter{\hbox{{\includegraphics[scale=#1]{#2}}}}}}

\usepackage{color}

\begin{document}

\author{Carlos~Fern\'andez-Gonz\'alez}
\affiliation{Departamento de F\'{\i}sica Interdisciplinar,
Universidad Nacional de Educaci\'on a Distancia (UNED), 28040 Madrid,
Spain}

\author{Roger~S.~K.~Mong}
\affiliation{Department of Physics and Astronomy, University of
Pittsburgh, Pittsburgh, PA 15260, United States}

\author{Olivier~Landon-Cardinal}
\affiliation{Department of Physics, McGill 
University, Montr\'eal (Qu\'ebec) H3A 2T8, Canada}

\author{David~P\'erez-Garc\'{\i}a}
\affiliation{Departamento de An\'alisis Matem\'atico \& IMI, Universidad
Complutense de Madrid, 28040 Madrid, Spain}
\affiliation{ICMAT, C/ Nicol\'as Cabrera, 
Campus de Cantoblanco, 28049 Madrid, Spain}

\author{Norbert~Schuch}
\affiliation{Max-Planck-Institute of Quantum Optics, Hans-Kopfermann-Str.\
1, D-85748 Garching, Germany}

\title{Constructing topological models by symmetrization: A PEPS study}

\begin{abstract}
Symmetrization of topologically ordered wavefunctions is a powerful method
for constructing new topological models.  Here, we study wavefunctions
obtained by symmetrizing quantum double models of a group $G$ in the
Projected Entangled Pair States (PEPS) formalism.  We show that
symmetrization naturally gives rise to a larger symmetry group $\tilde G$
which is always non-abelian. We prove that by symmetrizing on sufficiently
large blocks, one can always construct wavefunctions in the same phase as
the double model of $\tilde G$. In order to understand the effect of
symmetrization on smaller patches, we carry out numerical studies for the
toric code model, where we find strong evidence that symmetrizing on
individual spins gives rise to a critical model which is at the phase
transitions of two inequivalent toric codes, obtained by anyon
condensation from the double model of $\tilde G$.
\end{abstract}

\maketitle

\section{Introduction}

Topologically ordered states are exotic phases of matter which do not
exhibit conventional order, but are characterized by their global
entanglement pattern which leads to effects such as a topological ground
space degeneracy or excitations with unconventional statistics, so-called
anyons.  A question of particular interest is how to construct new
topologically ordered states from existing ones, with different and
especially more complex topological order. One such route is anyon
condensation~\cite{bais:anyon-condensation,%
eliens:diagrammatic-anyon-condensation,neupert:boson-condensation}, which
removes part of the anyons from the model and generally leads to a simpler
anyon theory. A different route, which has been particularly successful
for fractional quantum Hall systems, is the projective construction. Here,
one starts from two or more copies of an initial wavefunction and projects
them locally, such as onto singly-occupied sites or onto the symmetric
subspace, ideally yielding a wavefunction which exhibits more rich
physics.  This way, one can for instance construct Lauglin states from
non-interacting topological insulators, or non-abelian Read-Rezayi states
from an abelian Laughlin state~\cite{read:projective,cappelli:coset}.
Recently, symmetrization of multiple copies has also been applied to
quantum spin systems with non-chiral topological order, such as Kitaev's
toric code model or quantum doubles~\cite{kitaev:toriccode}, and it has
been found that such a construction can indeed give rise to wavefunctions
with non-abelian characteristics by starting from an abelian
model~\cite{paredes:sym-tcode-1,paredes:sym-tcode-2}.

Projected Entangled Pair States (PEPS) provide a framework for the local
description of entangled quantum spin systems~\cite{verstraete:mbc-peps},
and allow to exactly capture fixed point wavefunctions with non-chiral
topological order such as quantum
double~\cite{kitaev:toriccode,verstraete:comp-power-of-peps,schuch:peps-sym}
or string-net models~\cite{levin:stringnets,buerschaper:stringnet-peps,%
gu:stringnet-peps}. In this framework, global topological order can be
explained from a local symmetry in the entanglement degrees of freedom of
the underlying local tensor, which codifies the capability of the model to
exhibit topological order, and allows for the succinct description of its
ground state manifold and topological
excitations~\cite{schuch:peps-sym,sahinoglu:mpo-injectivity,%
bultinck:mpo-anyons}.  Using this description, one can efficiently check
numerically whether a model realizes the full topological model given by
the underlying symmetry, or rather a simpler model obtained from it by
condensation, and thus determine the topological phase of a given PEPS
wavefunction~\cite{schuch:topo-top,haegeman:shadows,%
duivenvoorden:anyon-condensation}.

\begin{figure}[b]
\includegraphics[width=\columnwidth]{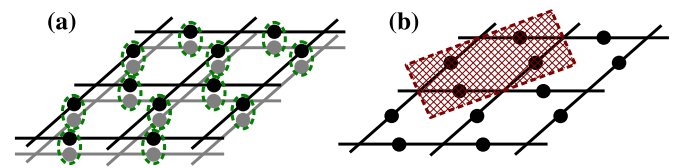}
\caption{
\label{fig:sym}
\textbf{(a)} General setup. We take two (or more) copies of a quantum
double model such as the toric code, where the spins are denoted by
dots, and project the copies locally onto the symmetric subspace,
indicated by the green ellipses. \textbf{(b)} Blocking scheme. After
blocking one plaquette as indicated, we can express any double model as a
$G$-isometric PEPS [Eq.~\eqref{eq:def-Giso-tensor}]. We also consider
wavefunctions which are symmetrized on such blocks; we will denote the
indicated block as a $1\times 1$ block, corresponding to the symmetrized
model SYM$_{1\times 1}$.
}
\end{figure}

In this paper, we apply PEPS to study models obtained by symmetrizing
topologically ordered wavefunctions, and to characterize their emergent
topological order.  Specifically, we consider models which are constructed
by taking two or more copies of a quantum double $D(G)$ with underlying
group $G$, such as the toric code, and projecting them locally onto the
symmetric subspace (Fig.~\ref{fig:sym}a). From a Hamiltonian point of
view, this requires the resulting model to be the ground state of a local
symmetrized Hamiltonian, and therefore to be \emph{locally}
indistinguishable from the symmetrized $D(G)$ wavefunction.

We show that within the PEPS framework, symmetrizing the wavefunction can
be understood through locally symmetrizing the corresponding tensors.
This induces an additional symmetry under \emph{locally} permuting the
copies, thus giving rise to tensors with a non-abelian symmetry group
$\tilde G:=(G\times G)\rtimes\mathbb Z_2$, or generalizations thereof,
which provide the right algebraic structure to characterize wavefunctions
which can locally be described as a symmetrized double $D(G)$.  We
analytically study the stucture of the symmetrized tensor and show that if
the symmetrization is carried out on sufficiently large patches
(Fig.~\ref{fig:sym}b), this always gives rise to a topological model in
the same phase as $D(\tilde G)$. In order to understand what happens when
we symmetrize on smaller regions, we additionally perform numerical
studies for the symmetrized toric code model $D(\mathbb Z_2)$, for which
$\tilde G=\mathrm{D}_4$, the dihedral group with $8$ elements, is
non-abelian.  We find that the symmetrized model is in the full
$D(\mathrm{D}_4)$ phase down to symmetrizations on blocks of $2\times 2$
plaquettes.  For symmetrization on smaller blocks, we find strong evidence
that the model is critical; in particular, symmetrization of single spins
gives rise to a model which sits at a phase transition between two
inequivalent toric code models, obtained from $D(\mathrm{D}_4)$ by two
different anyon condensations. Our work thus helps to clarify the nature
of wavefunctions obtained by symmetrizing topological spin models, and
demonstrates the power of PEPS to locally characterize topologically
ordered wavefunction and assess their structure by combining analytical
and numerical tools.

\section{Projected Entangled Pair States and topological order
\label{sec:peps-intro}}

\begin{figure}[b]
\centering
\includegraphics[scale=0.14]{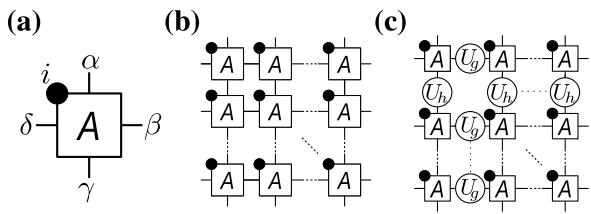}
\caption{
\label{fig:peps-def}
Construction of PEPS. \textbf{(a)} Five-index tensor $A$ with physical
index $i$ (dot) and virtual indices $\alpha,\dots,\delta$ (lines).
\textbf{(b)}~The PEPS is constructed by forming a 2D lattice and
contracting the connected virtual indices. \textbf{(c)}~Different ground
states on the torus can be parametrized by placing strings of symmetry
operations $U_g$, $U_h$, with $gh=hg$, around the torus.
}
\end{figure}

Let us start by introducing Projected Entangled Pair States (PEPS) and their
relation to topologically ordered states. We will consider a square
lattice on a torus of size $N_h\times N_v=:N$. A PEPS is constructed from
a $5$-index tensor $A^i_{\alpha\beta\gamma\delta}$ with \emph{physical
index} $i=1,\dots,d$ and \emph{virtual indices}
$\alpha,\dots,\delta=1,\dots,D$, with $D$ the \emph{bond dimension},
depicted in Fig.~\ref{fig:peps-def}a. The tensors are placed on the
lattice sites, and adjacent indices are contracted (i.e., identified and
summed over), indicated by connected lines in Fig.~\ref{fig:peps-def}b. We
are then left with an $N$-index tensor $c_{i_1\dots i_N}$ which defines
the PEPS wavefunction $\ket\psi = \sum c_{i_1\dots i_N}\ket{i_1,\dots,
i_N}$.

PEPS can exactly capture topological fixed point wavefunctions, such as
quantum double and string net models~\cite{verstraete:comp-power-of-peps,%
schuch:peps-sym, buerschaper:stringnet-peps,gu:stringnet-peps}.  In
particular, for the quantum double $D(G)$ of a finite group $G$, the PEPS
tensor $A$ describing a block of $2\times 2$ spins as in
Fig.~\ref{fig:sym}b, i.e., one plaquette, is up to local unitaries 
on the physical indices of the form
\begin{equation}
\label{eq:def-Giso-tensor}
\figg{0.3}{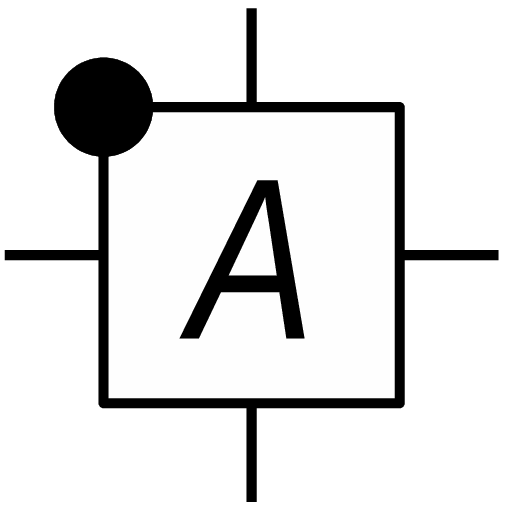}=\frac{1}{|G|}\sum_{g\in G}\figg{0.3}{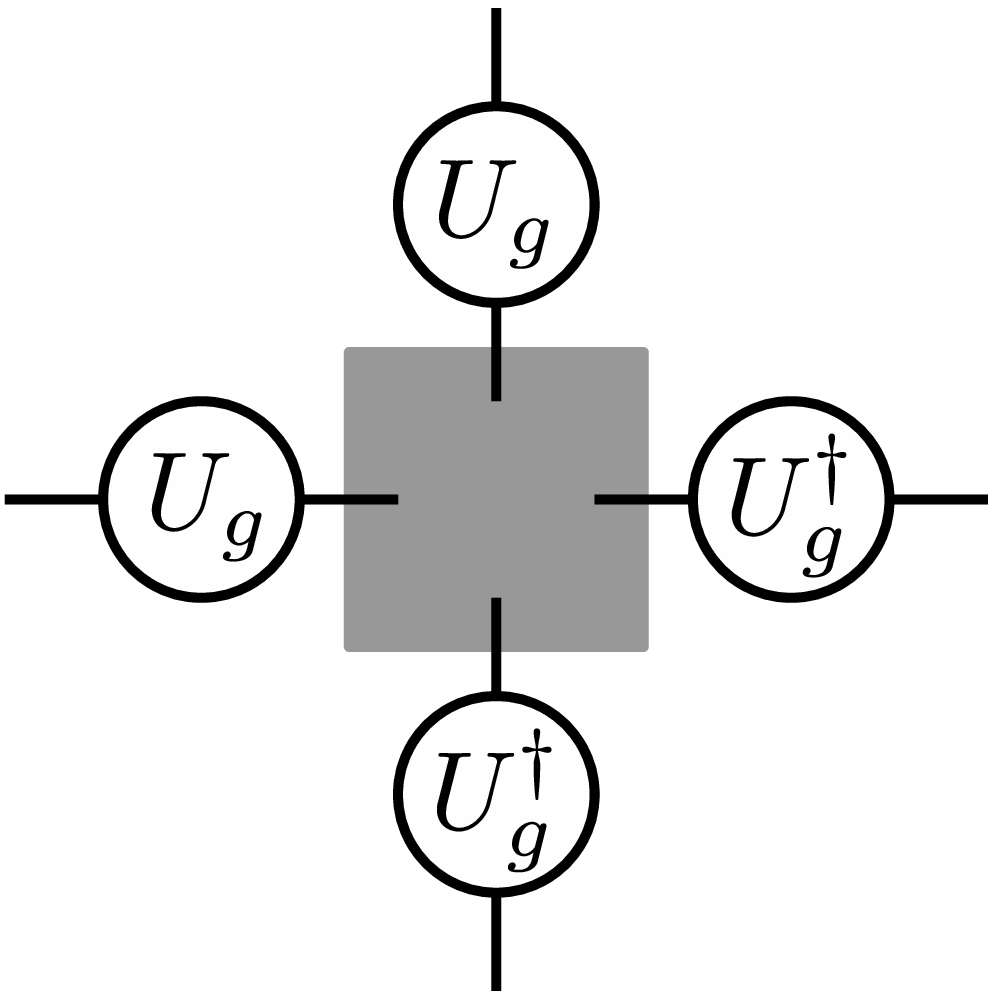}
\end{equation}
where on the r.h.s., the legs inside the shaded area jointly describe the
physical index, the $D\times D$ matrices $U_g=\sum_h \ket{gh}\bra{g}$ are
the regular representation of $G$ (thus $D=|G|$), and they act from right
to left and bottom to top. Differently speaking, if $A$ is interpreted as
a map $\mathcal P_A$ from the virtual to the physical indices, it is of
the form $\mathcal P_A=\tfrac{1}{|G|}\sum_g U_g\otimes U_g\otimes
\bar{U}_g\otimes \bar{U}_g$, the projector onto the invariant subspace. As
$\mathcal P_A$ is an isometry on the $G$-invariant subspace, these PEPS
are denoted as $G$-isometric. Note that
such an $A$ is invariant under applying $U_g$ (or
$U_g^\dagger$) to all virtual indices at the same time; this can in
particular be used to parametrize the ground state manifold of the model's
parent Hamiltonian by placing strings of $U_g$ ($U_h$) along the
horizontal (vertical) boundary (Fig.~\ref{fig:peps-def}c), which can be
freely moved (as long as $gh=hg$) due to the aforementioned
condition~\cite{schuch:peps-sym}.

We can modify a $G$-isometric tensor by adding deformations to the
physical system, $\mathcal P_{A'} = \Delta\cdot \mathcal P_A$; the resulting
tensor is still invariant under applying $U_g$ and thus allows for the
same ground space parametrization. As long as $\Delta$ is invertible, the
deformation can be ``kicked back'' onto the Hamiltonian, implying that the
ground space degeneracy in the finite volume remains
unchanged~\cite{schuch:peps-sym,schuch:rvb-kagome}; this might break down,
however, in the thermodynamic limit~\cite{schuch:topo-top}, as we will
also see later.  Of particular interest later on will be the case where
$\Delta=\Gamma^{\otimes 4}$ acts on all four physical sub-indices
independently, replacing $U_g$ by $X_g:=U_g\Gamma_g$, and $\Gamma$
commutes with $U_g$, i.e., it only changes the weight of each irrep
sector. 

An important observation is that the choice of $U_g$ (or $X_g$) is not unique,
and there are many different representations for a wavefunction constructed
from tensors of the form Eq.~(\ref{eq:def-Giso-tensor}) which are
equivalent up to local unitaries.  To this end, consider two adjacent
sites of a PEPS
constructed from such tensors,
\[
\frac{1}{|G|^2}\sum_{g,h\in G}
\figg{0.28}{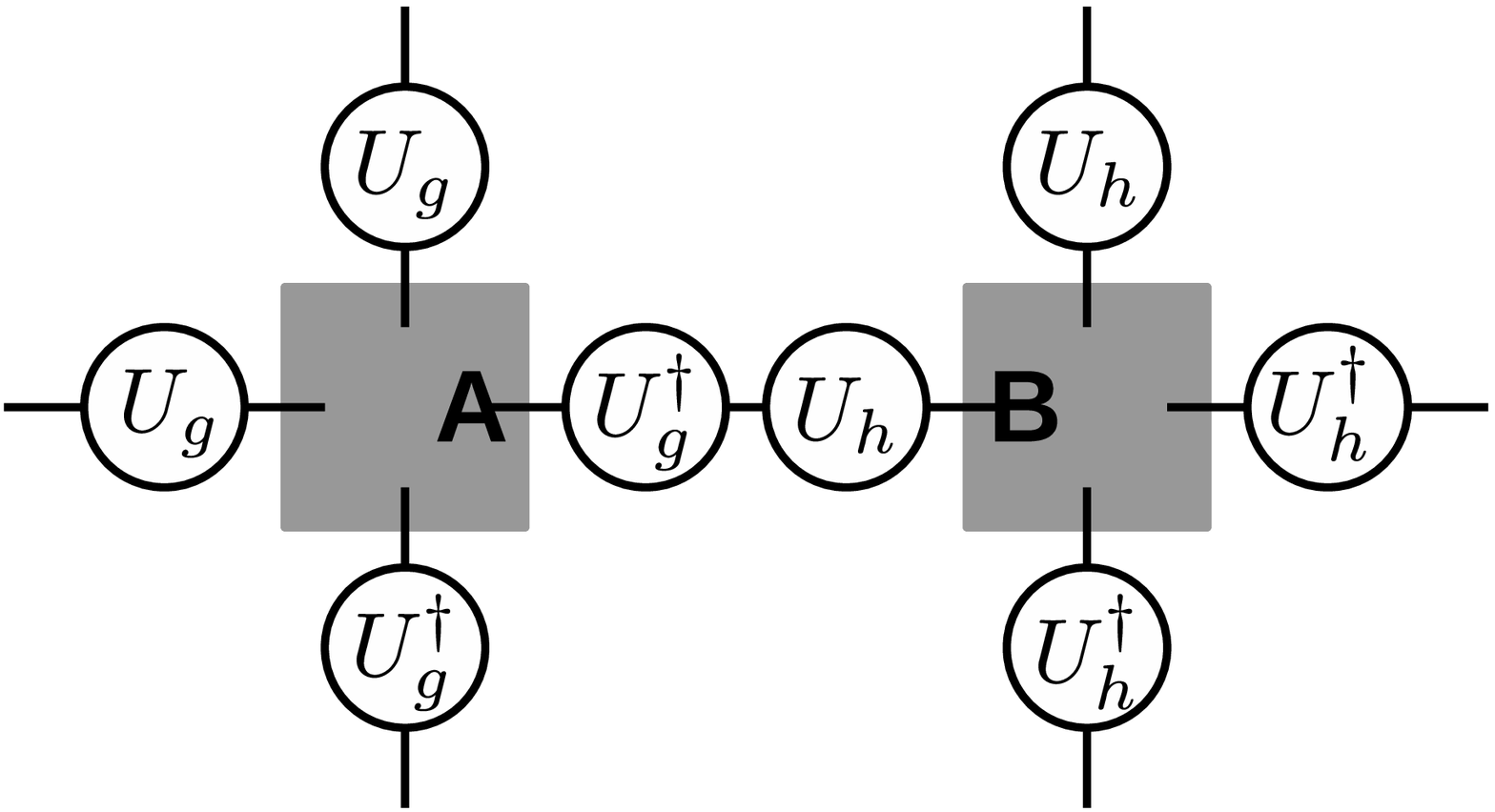}\ ,
\]
where either $U_g\equiv X_g$ or $U_g\equiv Y_g$, without any
assumption on the structure of $X_g$ and $Y_g$.  Then, the state between
the physical spins labelled $A$ and $B$ is of the form
$\ket{\chi_{g,h}(U)} = \sum_{ij}(U_g^\dagger
U^{\phantom\dagger}_h)_{ij}\ket{i,j}$, where $U=X,Y$.  However, as long as
the angles between the $\ket{\chi_{g,h}(U)}$ are preserved  -- this is, 
\begin{equation}
\label{eq:equiv-angles}
\mathrm{tr}[X_g^\dagger X_h^{\phantom\dagger} 
    X_{h'}^\dagger X_{g'}^{\phantom\dagger}] =
\mathrm{tr}[Y_g^\dagger Y_h^{\phantom\dagger} 
    Y_{h'}^\dagger Y_{g'}^{\phantom\dagger}]
\end{equation}
-- we can always find an isometry $T$ acting on $AB$ which maps
$T:\ket{\chi_{g,h}(X)}\mapsto \ket{\chi_{g,h}(Y)}$, and thus the
corresponding PEPS with $U_g\equiv X_g$ or $U_g\equiv Y_g$ are equivalent up to
local unitaries. An important special case of this equivalence is given by
the case where a unitary representation $X_g\equiv U_g=\bigoplus_\alpha
D^\alpha(g)\otimes \mathbb{I}_{m_\alpha}$  with irreducible
representations (irreps) $D^\alpha$ with multiplicity $m_\alpha$ is
replaced by $Y_g\equiv \Gamma \hat U_g$, where $\hat U_g = \bigoplus D^\alpha(g)$ is
multiplicity-free, and $\Gamma = \bigoplus m_\alpha^{1/4} \mathbb
I_{d_\alpha}$ (with $d_\alpha$ the dimension of the irrep $\alpha$)
adjusts the weight of the irreps~\cite{schuch:peps-sym}.

We thus see that all what matters when characterizing a $U_g$-invariant
tensor of the form \eqref{eq:def-Giso-tensor}, with $U_g=\bigoplus_\alpha
w_\alpha^{1/4}D^\alpha(g)\otimes \mathbb I_{m_\alpha}$, is the total weight
$w_\alpha m_\alpha$ of each irrep in $U_g$.  In particular, given a tensor
\eqref{eq:def-Giso-tensor} for which the relative weights of the irreps
of $U_g$ are sufficiently close to those in the regular representation,
and thus the deformation $\Delta$ required to relate the system to the
fixed point model will be sufficiently close to the identity, we can show
that the gap of the parent Hamiltonian will remain open, and the system
will be in the $D(G)$ phase~\cite{schuch:mps-phases}, as will be
discussed in more detail later on~\footnote{Note that
the converse is not true: There can be tensors of the form
Eq.~\eqref{eq:def-Giso-tensor} where the irrep weights in $U_g$ are very
far from the regular representation, but which nevertheless are in the
$D(G)$ phase; this is related to the fact that missing irreps can be
obtained by blocking (i.e., as irreps of $U_g\otimes U_g$). For instance,
this is the case for the model obtained from the $D(\mathrm{D}_4)$ model
by removing the trivial irrep from the regular representation.  }.

\section{Symmetrizing topological PEPS wavefunctions}

\subsection{Invariance of symmetrized wavefunctions}

Let us now consider what happens when we take two copies of a topological
PEPS with symmetry group $G$ and project them onto the symmetric subspace
on the physical system,
\[
\figg{0.3}{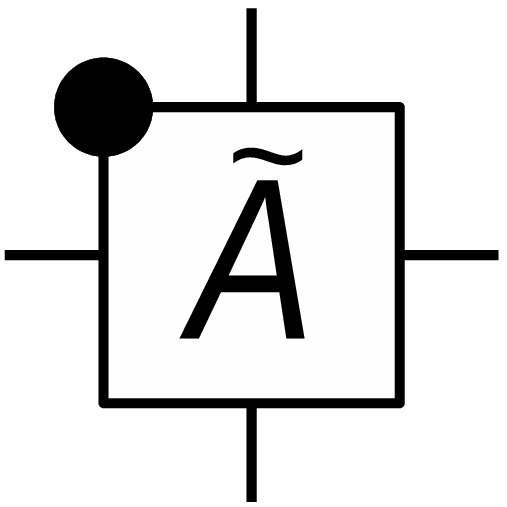}:=\quad\raisebox{0.6em}{\figg{0.3}{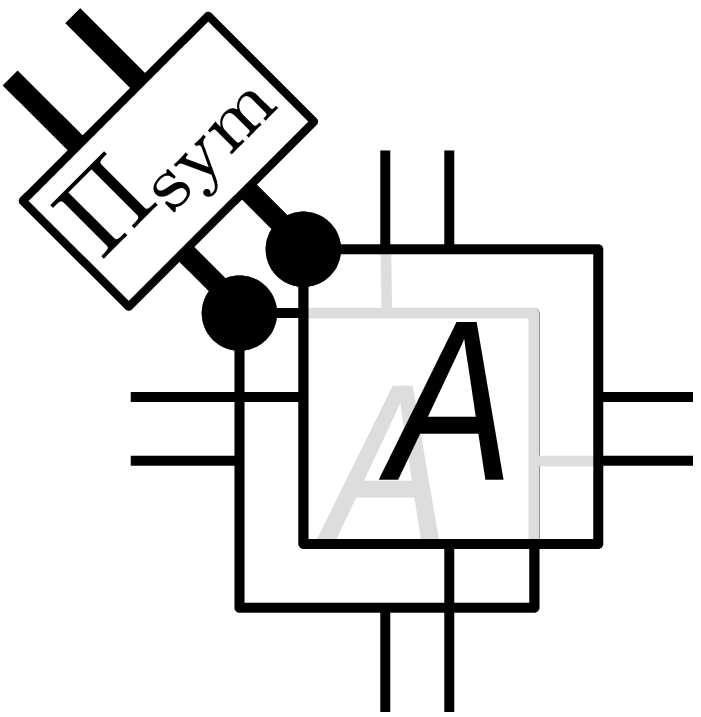}}
\]
with $\Pi_\mathrm{sym}$ the projection onto the symmetric subspace.
It is clear that the resulting PEPS will have a virtual $G\times G$
symmetry, obtained from the independent action $U_g\otimes U_h$ of the
original symmetry on the two copies.  However, there is also another
symmetry emerging: Since the symmetric subspace is invariant under
swapping the physical indices of the two tensors, and the latter is
equivalent to swapping the virtual indices between the copies,
\begin{equation}
\label{eq:phys-F-virt-F}
\raisebox{0.6em}{\figg{0.3}{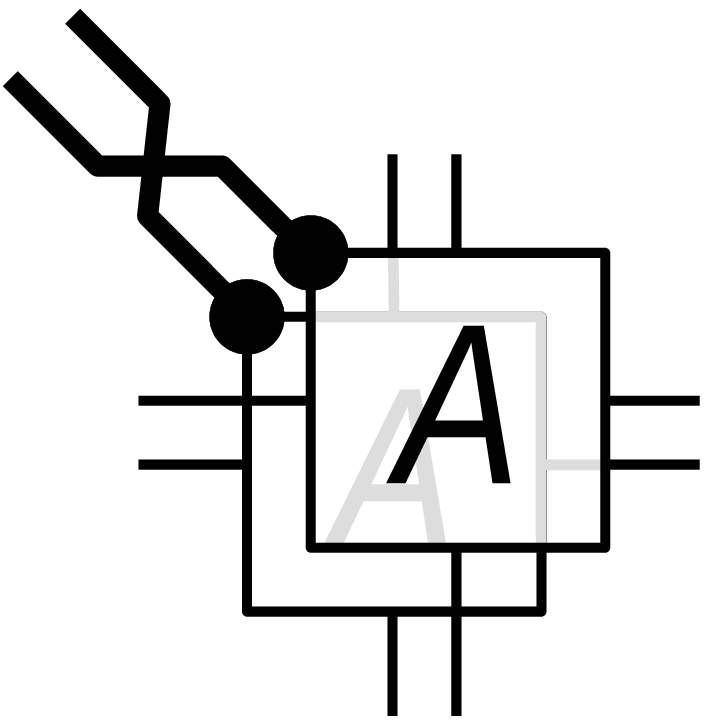}} \quad = 
\quad \raisebox{0.60em}{\figg{0.3}{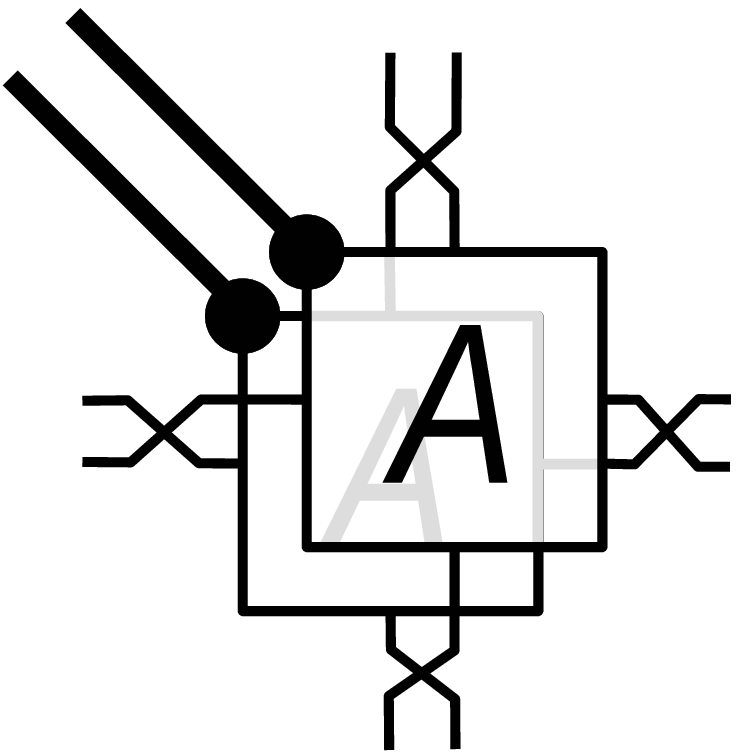}}\ ,
\end{equation}
we have that the symmetrized tensor is in addition invariant under the
``flip'' $\mathbb F$ which swaps the virtual spaces, $(U_g\otimes
U_h)\,\mathbb F=\mathbb F\,(U_h\otimes U_g)$.  The overall symmetry group
is thus $\tilde G:=(G\times G)\rtimes\mathbb Z_2$ which is generated by
the elements of the direct product $(g,h)\in G\times G$, together with the
semi-direct action $\mathbb F(g,h)\mathbb F=(h,g)$ of the non-trivial
element $\mathbb F \in \mathbb Z_2$; in particular, $\tilde G$ is
non-abelian also for abelian $G$. It is thus well possible that the
resulting symmetrized wavefunction has topological order described by
the non-abelian model $D(\tilde G)$. The corresponding ground states would
then be again of the form of Fig.~\ref{fig:peps-def}c, with $U_g$ a
representation of $\tilde G$, and are therefore obtained by either
symmetrizing two copies of the original ground states, or by additionally
inserting an $\mathbb F$ at either boundary which twists the two copies,
i.e., by wrapping one $D(G)$ model twice around the torus and
symmetrizing~(cf.~Ref.~\cite{repellin:Zk-groundstates}).

Let us study the tensor obtained by symmetrizing more closely. We have
$\Pi_\mathrm{sym}=\tfrac12(\mathbb I + \mathbb F)$, and for tensors
expressed as in Eq.~(\ref{eq:def-Giso-tensor}), $\mathbb F\equiv \mathbb
F^{\otimes 4}$ factorizes over the four physical indices.  By combining
the sum over $U_g$ and $U_{h}$ in the two copies of $A$ with the sum over
$\{\mathbb I,\mathbb F\}$ in $\Pi_\mathrm{sym}=\tfrac12(\mathbb I^{\otimes
4}+\mathbb F^{\otimes 4})$, we thus find that the symmetrized PEPS tensor
$\tilde A$ is of the form 
\begin{equation}
\label{eq:symtensor-Giso}
\figg{0.3}{Atilde}=\frac{1}{|\tilde G|}\sum_{\bm k\in \tilde G}\figg{0.3}{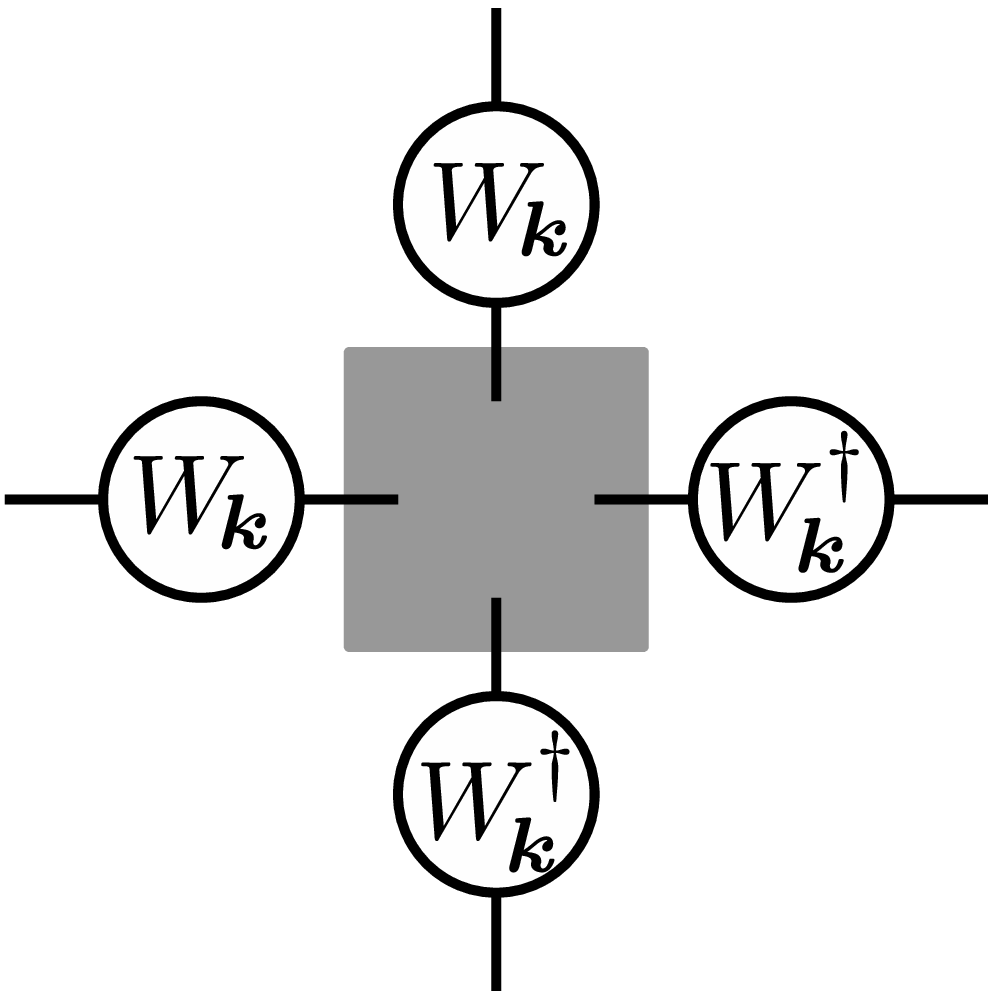}
\end{equation}
where 
\begin{align*}
W_{\bm k}&:=U_g\otimes U_h && \mbox{\ for\ }{\bm k}=(g,h)\in \tilde G\ ,
\\
W_{\bm k}&:=(U_g\otimes U_h) \mathbb F  && \mbox{\ for\ } 
    {\bm k}=(g,h)\mathbb F\in \tilde G
\end{align*}
forms a unitary representation of $\tilde G$.  Note that the fact that
$W_{\bm k}$ is a unitary representation immediately implies that $\mathcal
P_{\tilde A}$ is a projector.

\subsection{Representation structure of symmetrized tensor}

Let us now study the model obtained by symmetrizing more closely.  We will
in the following restrict to the case of an abelian group $G$, since in
this case, symmetrizing holds the promise to transform an abelian model
into a non-abelian one; for the general case,
cf.~Appendix~\ref{sec:app:sym-character}.  In order to see whether the
symmetrized tensor $\tilde A$ describes a topological model related to
$D(\tilde G)$, we need to study the irrep structure of $W_{\bm k}$:
if the relative weights of the different irreps are sufficiently close 
to the weights in the regular representation of $\tilde G$, this implies
that the model is in the same phase.

We start by splitting the regular representation $U_g$ into its
one-dimensional (1D) irreps, $U_g=\bigoplus_{\alpha=1}^n D^\alpha(g)$,
where each $D^\alpha$ is supported on a one-dimensional Hilbert space
$\mathcal H_\alpha$, $\bigoplus_{\alpha=1}^n \mathcal H_\alpha = \mathbb
C^d$.  This yields a corresponding decomposition
\[
U_g\otimes U_h=\bigoplus_{\alpha,\beta=1}^nD^\alpha(g)\otimes D^\beta(h)\ ,
\]
where $D^\alpha(g)\otimes D^\beta(h)$ acts on $\mathcal H_\alpha\otimes
\mathcal H_\beta$; $n=|G|$ is the order of the group. We now distinguish
two cases: For $\alpha=\beta$, we have that each $\mathcal H_\alpha\otimes
\mathcal H_\alpha$ is invariant both under
$U_g\otimes U_h$ and under $\mathbb F$, thus being a 1D irrep; as the
action of $(g,1)$ is different for each $\alpha$, the irreps are all
different. On the other hand, for $\alpha\ne\beta$, 
$\mathcal H_\alpha\otimes \mathcal H_\beta\oplus\mathcal H_\beta\otimes
\mathcal H_\alpha$ is again invariant under both $U_g\otimes U_h$ and
$\mathbb F$, and since $U_g\otimes U_h$ and $\mathbb F$ have
different eigenbases, this is a 2D irrep. There are at most
$\tfrac{n(n-1)}{2}$ such irreps. 

In order to check whether we have obtained all irreps, we can now use the
counting formula for the number of irreps, $\sum d_\alpha^2=|\tilde G|$.
From the preceding arguments, we find that for the irreps we found, $\sum'
d_\alpha^2 \le n\, 1^2+ \tfrac{n(n-1)}{2}\,2^2 = 2n^2 -n$, while $|\tilde
G|=2n^2$. We thus see that $W_{\bm k}$ for the symmetrized wavefunction is
even missing some irreps, and we therefore do not expect the resulting
wavefunction to be the phase of the double model $D(\tilde G)$.

\subsection{Blocking and symmetrization}

The reason for the missing representations seems related to the fact that
$\mathcal H_\alpha\otimes\mathcal H_\alpha$ is one-dimensional, and thus
$\mathbb F$ can only act trivially on it. A way to remedy this would be to
have irreps with higher multiplicities. To this end, we might try to
symmetrize larger blocks of the system, which correspondingly carry larger
representations.  We therefore introduce a new tensor $B$ obtained by
blocking $\ell\times \ell$ tensors $A$,
$$\figg{0.14}{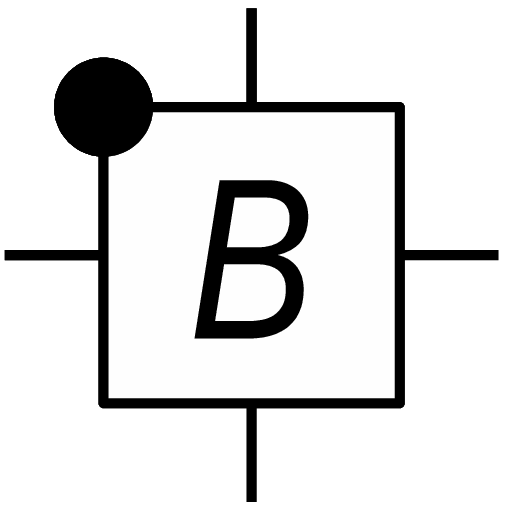}:=\left.\figg{0.14}{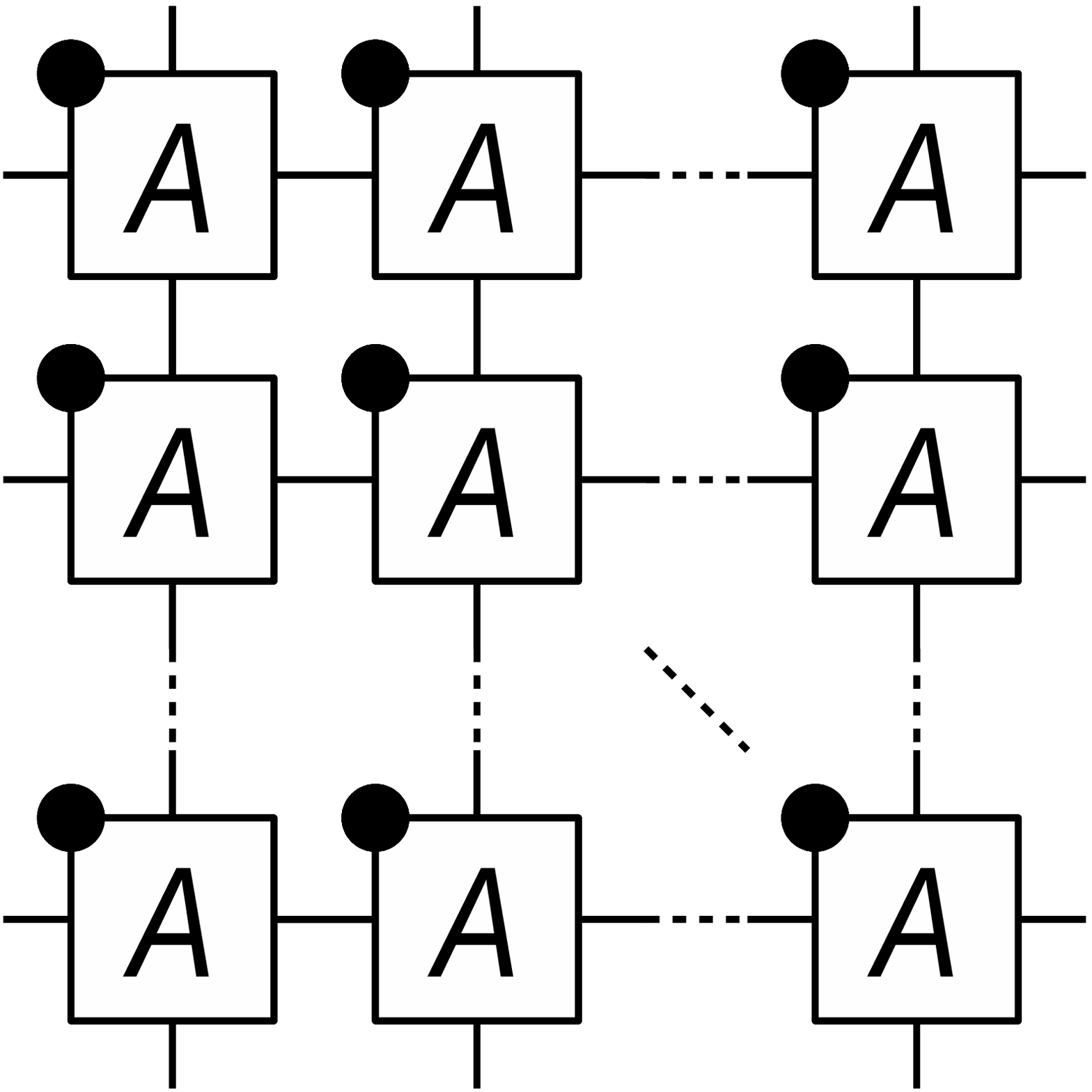}\right\}\ell\mathrm{\ times}$$ and
subsequently take two copies of $B$ rather than $A$ and project them onto
the symmetric subspace, yielding a tensor $\tilde B$. 

Up to a local isometry on the physical system and an additional
normalization factor $|G|^{-(\ell-1)}$, $B$ is of the form
Eq.~(\ref{eq:def-Giso-tensor}), but with $U_g$ replaced by
$V_g=U_g^{\otimes\ell}$, as shown in Ref.~\cite{schuch:peps-sym}. Since
$U_g^{\otimes \ell}\cong U_g\otimes \mathbb I_{n}^{\otimes \ell-1}$, we can
decompose $V_g$ as
\[
V_g=\bigoplus_{\alpha=1}^n D^\alpha(g)\otimes{\mathbb I}_{\bar n}
\]
where $\bar n:=n^{\ell-1}$, and $D^\alpha(g)\otimes{\mathbb
I}_{\bar n}$ is supported on an $\bar n$-dimensional subspace $\mathcal
K_\alpha$ of $(\mathbb C^n)^{\otimes\ell} = \bigoplus_{\alpha=1}^n 
\mathcal K_\alpha$. As before, the symmetrized tensor $\tilde B$ is of the
form Eq.~(\ref{eq:symtensor-Giso}), but now with a representation
$W'_{(g,h)}:=V_g\otimes V_h$ and $W'_{(g,h){\mathbb F}} := (V_g\otimes
V_h){\mathbb F}$.

Just as before, for each $\alpha=1,\dots,n$ the subspace $\mathcal
K_\alpha\otimes \mathcal K_\alpha$ is invariant both under all $V_g\otimes
V_h$ and $\mathbb F$.
However, we can now further decompose the corresponding subrepresentation.
The subrepresentation is
\[
\begin{array}{rcl}
(g,h)& \mapsto &D^\alpha(g)\otimes D^\alpha(h)\otimes {\mathbb I}_{\mathcal K_\alpha\otimes \mathcal K_\alpha}\\
(g,h){\mathbb F} & \mapsto & D^\alpha(g)\otimes D^\alpha(h)\otimes {\mathbb F}_{\mathcal K_\alpha\otimes \mathcal K_\alpha}
\end{array}
\]
and all these matrices commute since $D^\alpha(g)\otimes D^\alpha(h)$ is a
scalar, i.e., they are proportional to $\mathbb I$ or $\mathbb F$.
Therefore, it can be further split into different irreps by considering the
eigenspaces of $\mathbb F$, namely the symmetric subspace $\mathcal
S(\mathcal K_\alpha\otimes \mathcal K_\alpha)$ and the antisymmetric
subspace $\mathcal A(\mathcal K_\alpha\otimes \mathcal K_\alpha)$ with
 eigenvalues $+1$ and $-1$, respectively.  On these two
subspaces, we have subrepresentations
$$
\begin{array}{rcl}
(g,h)\phantom{\mathbb{F}}& \mapsto &\phantom{-}D^\alpha(g)\otimes D^\alpha(h)
	\otimes {\mathbb I}_{\mathcal S(\mathcal K_\alpha\otimes \mathcal K_\alpha)}\\
(g,h){\mathbb F} & \mapsto & \phantom{-}D^\alpha(g)\otimes D^\alpha(h)\otimes 
	{\mathbb I}_{\mathcal S(\mathcal K_\alpha\otimes \mathcal K_\alpha)}
\end{array}$$
and
$$
\begin{array}{rcl}
(g,h)\phantom{\mathbb{F}}& \mapsto &\phantom{-}D^\alpha(g)\otimes D^\alpha(h)\otimes {\mathbb I}_{\mathcal A(\mathcal K_\alpha\otimes \mathcal K_\alpha)}\\
(g,h){\mathbb F} & \mapsto & -D^\alpha(g)\otimes D^\alpha(h)\otimes
{\mathbb I}_{\mathcal A(\mathcal K_\alpha\otimes \mathcal K_\alpha)}\ ,
\end{array}
$$
respectively.  Each of these is a 1D irrep with multiplicity
$\mathrm{dim}\,\mathcal S(\mathcal K_\alpha\otimes \mathcal K_\alpha)=\bar n(\bar
n+1)/2$ and 
$\mathrm{dim}\,\mathcal A(\mathcal K_\alpha\otimes \mathcal K_\alpha)=\bar n(\bar
n-1)/2$, respectively, and there are $n$ of each kind (one for each
$\alpha$). Clearly, all of these $2n$ irreps are distinct, since they act
differently on $(g,1)$ and/or $(1,1)\mathbb F$.

Let us now turn to the subspaces $\mathcal K_\alpha\otimes \mathcal
K_\beta \oplus \mathcal K_\beta\otimes \mathcal K_\alpha$ ($\alpha\ne
\beta$). Each of them is again invariant under both $V_g\otimes V_h$ and
$\mathbb F$. To decompose them further, consider a basis
$\{e_1,\dots,e_{\bar n}\}$ of $\mathcal K_\alpha$ and a basis
$\{e'_1,\dots,e'_{\bar n}\}$ of $\mathcal K_\beta$.  For each
$s,t\in\{1,\dots,\bar n\}$, the subspace $\mathrm{span}\{e_s\otimes
e'_t,e'_t\otimes e_s\}$ is still invariant under $V_g\otimes V_h$ and
$\mathbb F$. This space corresponds to a 2D irrep, since for the
corresponding subrepresentation, the elements $D^\alpha(g)\otimes
D^\beta(h)\oplus D^\beta(g)\otimes D^\alpha(h)$ are diagonal in the basis
$\{e_s\otimes e'_t,e'_t\otimes e_s\}$, while $\mathbb F$ is diagonal in
the basis $\{e_s\otimes e'_t\pm e'_t\otimes e_s\}$.  We thus obtain $\bar
n^2$ copies of a 2D irrep for each pair $\{\alpha,\beta\}$ with
$\alpha\ne\beta$. Again, we get different irreps for each such pair since
they act differently on $(g,1)$.  Thus, we obtain in total $n(n-1)/2$
different 2D irreps, each with multiplicity $\bar n^2$.

In total, we have $2n$ distinct 1D irreps and $n(n-1)/2$ distinct 2D
irreps, and thus $\sum d_\alpha^2=2n+4n(n-1)/2=2n^2=|\tilde G|$: We have
found that by blocking at least $2\times 2$ sites, we obtain a symmetrized
tensor $\tilde B$ of the form Eq.~(\ref{eq:symtensor-Giso}), where $W_{\bm
k}$ carries all irreps of $\tilde G$.

Of course, this still does not imply that the symmetrized PEPS described
by $\tilde B$ is in the same phase as the $D(\tilde G)$ quantum double.
However, we know that we can continuously deform $\tilde B$ by acting with
some $\Gamma$ on each of the four physical indices in
Eq.~\eqref{eq:symtensor-Giso} which changes the weights of the irreps, in a
way which allows us to continuously deform $\tilde B$ to a tensor which
corresponds to the fixed point wavefunction of the quantum double
$D(\tilde G)$. Such a smooth deformation of the tensor can be ``kicked
back'' to a deformation of the parent Hamiltonian $H=\sum
h_i$~\cite{schuch:rvb-kagome} such that the $h_i$ change continuously as
well.  While this clearly does not imply that the system is in the same
phase, it will be so in the vicinity of the fixed point wavefunction,
i.e., if the deformation $\Delta=\Gamma^{\otimes 4}$ is sufficiently close
to the identity; in particular, one can derive bounds on the
deformation~\cite{schuch:mps-phases} for which one can prove that the
deformed Hamiltonian is connected to the gapped fixed point Hamiltonian
through a gapped path, thereby guaranteeing that the system is in the
$D(\tilde G)$ topological phase.  For the case under consideration, the
relative weights $\tfrac{\bar n(\bar n+1)}{2}:\tfrac{\bar n(\bar
n-1)}{2}:\bar n^2$ of the irreps need to be changed to $1:1:2$ (modulo
normalization) for each of the four physical indices; the ratio of
smallest and largest eigenvalue of $\Delta$ is thus $\rho = \tfrac{\bar
n-1}{\bar n+1}$ (as each deformation $\Gamma$ only carries the fourth root
of the multiplicity, cf.~Sec.~\ref{sec:peps-intro}). A straightforward
application of Appendix E of Ref.~\cite{schuch:mps-phases}, where it is
shown that a deformation of up to $\rho\ge \rho_0\approx0.967$ does not
close the gap, yields that $\bar n \ge 2/(1-\rho_0)-1 \approx 59.6$. 
We thus find that for a model with $|G|=n=2$, symmetrizing over a block of
size $\ell\ge7$ gives rise to  a model in the $D(\tilde G)$ phase, while
for $|G|\ge59$, $\ell=2$ is sufficient.

\section{Numerical study}

As we have seen, symmetrizing a $G$-isometric PEPS wavefunction on a
sufficiently large block gives a system which is in the phase of the
$D(\tilde G)$ topological model.  However, what happens if we symmetrize
on a smaller scale, such as on single tensors, or even
on the level of a single site in the toric code?  

In order to understand this question, we can study smooth interpolations
between a model of interest and a point which we understand analytically.
As long as the interpolation is described by a smooth and invertible map
acting on
the physical index, the deformation corresponds to a smooth deformation
of the corresponding parent Hamiltonian, and we can investigate whether
along such an interpolation the system undergoes a phase transition.
A convenient way to carry out such interpolations between tensors symmetrized
on arbitrary block size is to use the fact that by local unitaries
they can be transformed into a form where each irrep only appears once,
but weighted with a diagonal matrix which accounts for the multiplicity of
the irreps, as explained around Eq.~(\ref{eq:equiv-angles}), and
interpolate the corresponding weights. In
particular, this allows us to interpolate all the way from the fixed point
wavefunction through a model symmetrized on a $2\times 2$ block down to
the model symmetrized on a single plaquette.  Note that we can
equivalently understand this as a procedure where we start from a model
with a $2\times 2$ plaquette unit cell which is locally equivalent to the
RG fixed point wavefunction $D(\tilde G)$, from which we interpolate to
the $2\times 2$ plaquette symmetrized wavefunction SYM$_{2\times 2}$ and
then all the way to a wavefunction SYM$_{1\times 1}$ which is equivalent
up to local unitaries to the wavefunction symmetrized on $1\times 1$
blocks, cf.~Fig.~\ref{fig:sym}b.  (Note however that this is not the same
as interpolating from a $2\times 2$ symmetrized block to  a block of
$2\times 2$ tensors each of which has been individually symmetrized.)

\begin{figure}[t]
\includegraphics[width=\columnwidth]{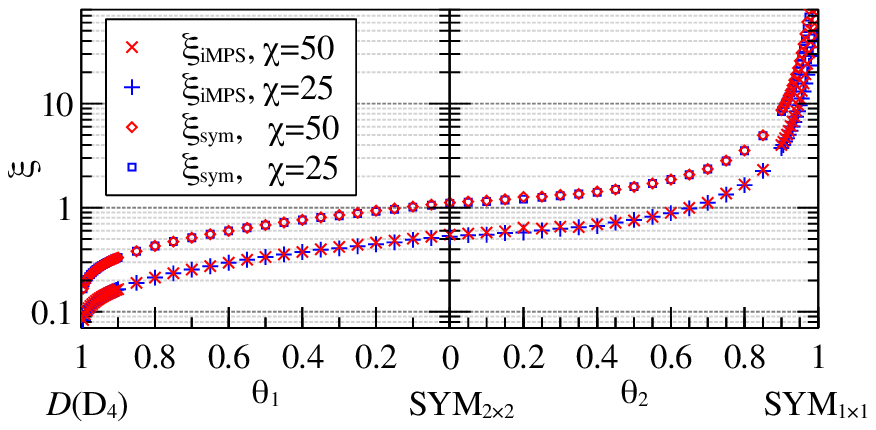}
\caption{
\label{fig:2x2to1x1}
Correlation length $\xi$ for the interpolation from the fixed point
$D(\mathrm{D}_4)$ through the model symmetrized on $2\times 2$ plaquettes
(SYM$_{2\times 2}$) to the model symmetrized on $1\times 1$ plaquette
(SYM$_{1\times 1}$), cf.~Fig.~\ref{fig:sym}b; we find that the correlation
length only diverges when getting close to SYM$_{1\times 1}$. The
interpolation was realized by linearly interpolating between the irrep
weights $w_\alpha$ of the $D(\mathrm{D}_4)$ and SYM$_{2\times 2}$ model
and those of the SYM$_{2\times 2}$ and SYM$_{1\times 1}$ model, specified
by the parameters $\theta_1$ and $\theta_2$, respectively.
We have approximated the fixed point of
the transfer operator by an iMPS with bond dimensions $\chi=25,50$ using up
to $200$ iterations, and extracted the correlation length from the fixed
point MPS in two ways: $\xi_\mathrm{iMPS}$ corresponds to the correlation
length of the iMPS itself [i.e., $-1/\log(\lambda_2/\lambda_1)$, with
$\lambda_{1,2}$ the leading eigenvectors of its transfer matrix], which
captures correlations between topologically trivial excitations as well as
purely electric excitations (which do not carry a string in the PEPS
representation~\cite{schuch:peps-sym}). $\xi_{\mathrm{sym}}$ is the
largest length scale set by anyon-anyon correlations with non-trivial flux
(determined from the largest eigenvalue of all mixed transfer operators with
a flux string $W_{\bm k}\otimes \bar{W}_{\bm k'}$
inbetween~\cite{haegeman:shadows,duivenvoorden:anyon-condensation}). 
Here, the largest length scale is given by $\bm k\in C_4$, $\bm k'\in
C_1$, corresponding to the mass gap of an anyon with flux in $C_4$ (see
Appendix~\ref{sec:app:anyons}). Note that further information on the
anyons could be extracted by labelling the eigenvectors by irreps.
}
\end{figure}

We have studied the corresponding interpolation for the symmetrized toric
code model, with the result for the correlation length shown in
Fig.~\ref{fig:2x2to1x1}: We find that the correlation length stays bounded
throughout the interpolation and only diverges at SYM$_{1\times 1}$,
demonstrating that the system is in the full $D(\tilde G)$ topological
phase all the way until the SYM$_{1\times 1}$ point. Note, however, that
the Hamiltonian does not change continuously at SYM$_{1\times 1}$, as
irreps vanish (though one can define a continuous but gapless
\emph{uncle
Hamiltonian}~\cite{fernandez:1d-uncle,fernandez-gonzalez:uncle-long},
and a parent Hamiltonian defined on larger patches might still be
continuous), and
while the results demonstrate that SYM$_{1\times 1}$ has diverging
correlations, the implications about the phase diagram should be taken
with care.

We have also considered the interpolation between SYM$_{1\times 1}$ and
SYM$_{\nicefrac12\times\nicefrac12}$, the point where we have symmetrized
the tensors on the level of individual spins in the toric code model
(Fig.~\ref{fig:sym}a), and have found strong evidence that the correlation
length diverges all the way thoughout the interpolation.  In order to
better understand the strucure of the symmetrized wavefunctions
SYM$_{\nicefrac12\times\nicefrac12}$ and SYM$_{1\times 1}$, we have
therefore considered a different interpolation which allows us to connect
these models with known gapped phases.

\begin{figure}[t]
\includegraphics{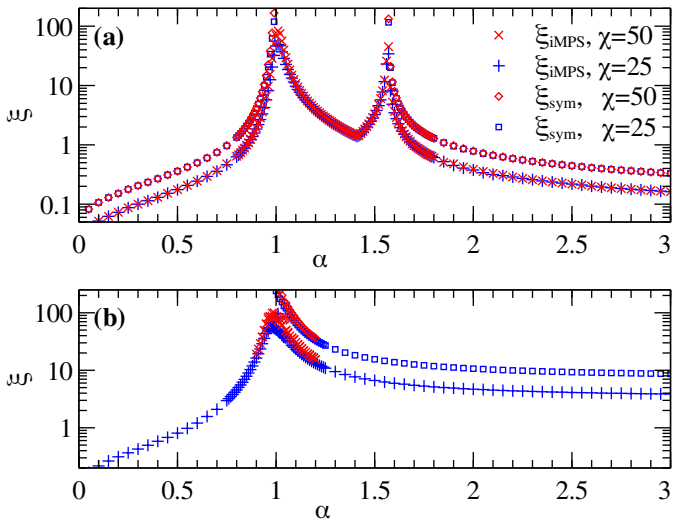}
\caption{
\label{fig:p5xp5-alpha}
\textbf{(a)} Correlation length for the $\alpha$-interpolation
[Eq.~\eqref{eq:alpha-interpol-p5}] for the toric code symmetrized on a
single spin; the symmetrized toric code point is at $\alpha=1$.  We find
three phases: two inequivalent toric code phases for $\alpha\lesssim 1$ and
$1\lesssim\alpha\lesssim 1.57$, and a trivial phase for $\alpha\gtrsim1.57$.
\textbf{(b)} Corresponding interpolation for SYM$_{1\times 1}$; we find a
toric code phase on the left and a $D(\mathrm{D}_4)$ phase on the right.
Cf.~Fig.~\ref{fig:2x2to1x1} for the numerical method and the meaning of
$\xi_\mathrm{iMPS}$ and $\xi_\mathrm{sym}$. In panel (a), in both the
small $\alpha$ and large $\alpha$ phase $\xi_\mathrm{sym}$ corresponds to
$\bm k=\bm k'\in C_3$ and gives thus the confinement length of particles
with flux in $C_3$; in panel (b), $\xi_\mathrm{sym}$ is again the mass gap of
$C_4$, as in Fig.~\ref{fig:2x2to1x1}.
}
\end{figure}

Let us first consider SYM$_{\nicefrac12\times\nicefrac12}$, which is
obtained by applying $\Pi_\mathrm{sym}=\tfrac12(\mathbb I+\mathbb F)$ to
each pair of spins in the two copies individually (with the spins in the
conventional loop-gas basis for the toric code).  We can now construct a
one-parameter family by replacing $\Pi_\mathrm{sym}$ with
\begin{equation}
\label{eq:alpha-interpol-p5}
\Pi(\alpha)=\ket{0,0}\bra{0,0}+\ket{1,1}\bra{1,1} +
\alpha\,\ket{\psi^+}\bra{\psi^+}\ ,
\end{equation}
where $\ket{\psi^+} = \tfrac{1}{\sqrt{2}}(\ket{0,1}+\ket{1,0})$. 
At $\alpha=0$, the projection locks the spins in the two copies to be
identical, and the resulting phase is unitarily equivalent to a single
copy of the toric code, for $\alpha=\infty$, the system becomes a trivial
product state, and for $\alpha=1$, we obtain the
SYM$_{\nicefrac12\times\nicefrac12}$ model.
Fig.~\ref{fig:p5xp5-alpha}a shows the correlation length (including
correlations between pairs of anyons) along this interpolation, which
gives strong indication for two phase transitions, one around
$\alpha\approx1$ and another one around $\alpha\approx1.57$. We
know that the phase on the very left is a toric code phase, and the one on
the very right is a trivial phase; but what about the intermediate phase?

In order to understand this phase, we can use the parametrization of the
ground space manifold of the deformed model in terms of the full symmetry
$W_{\bm k}$ of the $D(\tilde G)$ model, by placing string of symmetry
actions $W_{\bm k}$ when closing the
boundaries~\cite{schuch:peps-sym,schuch:topo-top},
cf.~Fig.~\ref{fig:peps-def}c.  This allows us to study the behavior of the
ground states labelled by the different particle types of the $D(\tilde
G)$ model. Here, we are interest in two types of information: First, what
is the norm of a ground state labelled by a non-trivial anyon relative to
the one labelled by the vacuum particle, and second, what is the
normalization of a ground state relative to the trivial sector?
Together, this allows us to understand
the ground state manifold in terms of condensation of anyons: If a ground
state becomes identical to the trivial ground state in the thermodynamic
limit, this implies that the corresponding anyon has condensed into the
ground state; and correspondingly, if the norm of a ground state is
vanishing, the corresponding anyon has become
confined~\cite{bais:anyon-condensation,schuch:topo-top,haegeman:shadows,%
duivenvoorden:anyon-condensation}. 

By applying this framework, we find that the intermediate phase around
$\alpha\approx 1.4$ is a toric code phase, just as the small $\alpha$
phase.  However, we also find that the two toric code phases are obtained
by condensing different particles into the vacuum, and one can therefore
indeed encounter a phase transition between them.  Let us note that using
the same analysis, we can verify that the phase around $\alpha=3$ is
indeed the trivial phase, as for $\alpha=\infty$.  Details on the method
and the analysis are given in Appendix~\ref{sec:app:anyons}.

We have also applied a similar analysis to SYM$_{1\times 1}$, the model
symmetrized on one plaquette, where we have replaced $\Pi_\mathrm{sym}$
(which now acts on  $4$ spins in the original toric code) by
\begin{equation}
\label{eq:alpha-interpol-1}
\Pi(\alpha) = \sum_i \ket{i,i}\bra{i,i} + \frac{\alpha}{2} 
    \sum_{i>j} \big[\ket{i,j} + \ket{j,i}\big]\,
		\big[\bra{i,j}+\bra{j,i}\big]\ .
\end{equation}
Note, however, that there are two shortcomings: First, the resulting model
is basis-dependent.  While for the single-site symmetrization, both
``natural'' bases for the toric code (which are related by a Hadamard
transformation) give the same state, this is no longer the case.
Here, we have chosen the $\ket{\pm}$ basis with the tensor as in Eq.~(7.5)
of Ref.~\cite{schuch:peps-sym}, since the dual choice gave
a very large correlation length even for $\alpha=\infty$.
Second, while for $\alpha=0$, the model is again the toric code, we cannot
analytically understand the structure of the model for $\alpha=\infty$,
and indeed, we find that it is not a fixed point wavefunction and exhibits
a finite correlation length $\xi\approx 3.4$.
Fig.~\ref{fig:p5xp5-alpha}b shows the result for the correlation length
along the interpolation; we find that the model again undergoes a phase
transition around $\alpha\approx 1$. By performing a similar analysis on
the ground state as before, we find that the small $\alpha$ phase is a
toric code phase, while the large $\alpha$ phase indeed realizes the
$D(\mathrm{D}_4)$ model, reconfirming that
SYM$_{\nicefrac{1}{2}\times\nicefrac{1}{2}}$ is at the boundary of the
$D(\mathrm{D}_4)$ phase.

\section{Conclusion and discussion}

We have analyzed the topological nature of wavefunctions obtained by
symmetrizing topologically ordered states.  We have shown that
symmetrizing a $G$-isometric PEPS, corresponding to a quantum double
$D(G)$, naturally gives rise to a symmetry $\tilde G=(G\times G)\rtimes
\mathbb Z_2$, with $\mathbb Z_2$ acting by permuting the components in the
tensor product.  While this gives the resulting wavefunction the
possibility to display $D(\tilde G)$ topological order, it can also
exhibit a simpler anyon theory obtained from $D(\tilde G)$ by
condensation. We were able to show that by symmetrizing on sufficiently
large blocks, one can always ensure that the resulting model is in the
$D(\tilde G)$ phase. The effect of symmetrization on smaller patches can
be analyzed numerically.  For the toric code $D(\mathbb Z_2)$, where
$\tilde G=\mathrm{D}_4$, we have found that the symmetrized model remains
in the full $D(\mathrm{D}_4)$ phase down to symmetrization on $2\times 2$
plaquettes.  For symmetrization on smaller blocks, the model appears to be
critical, sitting at a phase transition between inequivalent gapped
topological phases.

While we perfomed our analysis for the case of abelian groups $G$ and two
copies, it can immediately be generalized to the non-abelian case and
$k>2$ copies, in which case $\tilde G=G^{\times k}\rtimes \mathrm S_k$
(also known as the \emph{wreath product} of $G$ with $\mathrm S_k$).  Just
as before, one can show that the multiplicities of the irreps obtained by
symmetrizing $k$ copies of $U_g^{\otimes\ell}$, with $U_g$ the regular
representation,
will approach the correct ratio as $\ell$ grows, see
Appendix~\ref{sec:app:sym-character}.  Clearly, a similar analysis can
also be applied to string-net models, where the symmetry of the tensor is
itself described by a Matrix Product Operator
(MPO)~\cite{sahinoglu:mpo-injectivity}, on which we can define a
semidirect action of the flip just the same way.

One might wonder what happens when we symmetrize a trivial state,
$G=\{1\}$. In that case, $\tilde G=\mathrm S_k$, which can indeed support
topologically ordered states. However, since $n=|G|=1$, it will be
impossible to reach a regime where the symmetrized
wavefunction has all irreps by blocking when starting from the regular representation
of $G$.  This can be overcome by instead starting from, e.g., a
``trivial'' $D=2$ PEPS with maximally entangled bonds, with trivial group
action $U_1=\openone_D$; just as before, symmetrizing over sufficiently
large blocks (or using sufficiently large $D$) yields a model in the
$D(\mathrm{S}_k)$ phase (see Appendix~\ref{sec:app:sym-character}) which
for $k\ge3$ is again non-abelian.  While it might sound surprising that
symmetrizing a product state can give rise to topological order, note that
we symmetrize in a partition different from the one in which the state is
a product.  Also observe that since any group $G$ can be embedded in
$\mathrm{S}_{|G|}$, this allows to obtain any double model (such as one
universal for quantum computation) by symmetrizing a product state.  

An interesting perspective on symmetrized wavefunctions and their
excitations is in terms of lattice defects in topological
wavefunctions~\cite{bombin:defects,kitaev:gapped-boundaries}, and more
specifically so-called genons~\cite{barkeshli:genons} -- endpoints of
lattice defects in (non-symmetrized) multi-layer systems which allow
anyons to traverse between layers -- which can exhibit non-abelian
statistics even for abelian anyon models.  The genons are connected by
strings describing domain walls corresponding to the physical permutation
symmetry Eq.~\eqref{eq:phys-F-virt-F}, and are therefore confined. The
projection onto the symmetric subspace gauges this symmetry, promoting it
to a purely virtual symmetry on the entanglement degrees of freedom, and
thus transforms the confined lattice defects into potentially deconfined
anyonic excitations~\cite{barkeshli:u1-times-u1-rtimes-z2}.

\vspace*{-0.3cm}

\acknowledgements

We acknowledge helpful conversations with 
A.~Bernevig, 
A.~Essin,
G.~Evenbly,
J.~Haegeman, 
M.~Iqbal, 
B.~Paredes, and
B.~Yoshida.

CFG acknowledges support by the  Puentes Internacionales grant from UNED.
OLC is partially supported by Fonds de Recherche Quebec -- Nature et
Technologies and the Natural Sciences and Engineering Research Council of
Canada (NSERC).
DPG acknowledges support from MINECO (grant MTM2014-54240-P and ICMAT
Severo Ochoa project SEV-2015-0554) and Comunidad de Madrid (grant
QUITEMAD+-CM, ref.\ S2013/ICE-2801), from the John Templeton Foundation
through grant \#48322, and the European Research Council (ERC) under the
European Union's Horizon 2020 research and innovation programme (grant
agreement No 648913).
NS is supported by the Euroean Research Council (ERC) under grant no.\
636201 WASCOSYS, and the J\"ulich-Aachen Research Alliance (JARA) through
JARA-HPC grant jara0092. Part of this work was completed at the Aspen
Center for Physics, which is supported by National Science Foundation
grant PHY-1066293.
RSKM and OLC acknowledge funding provided by the Institute for Quantum
Information and Matter, an NSF Physics Frontiers Center (NSF Grants
PHY-1125565 and PHY-0803371) with support of the Gordon and Betty Moore
Foundation (GBMF-12500028), where part of this work was carried out.

\appendix

\section{Symmetrization of general groups
\label{sec:app:sym-character}}

In this appendix we consider the general case in which we  symmetrize $k$
copies of a $G$-isometric model, and show that the resulting representation
converges to multiple copies of the regular one under blocking.

Consider a finite group $G$ with regular representation $U_g$, and the
symmetric group $\mathrm{S}_k$ (or a subgroup thereof). Let $\tilde
G:=(G^{\times k})\rtimes \mathrm{S}_k$, where $\mathrm{S}_k$ acts by
permuting the $k$-fold product $G^{\times k}$.  (This is also known as the
wreath product $G\wr \mathrm{S}_k$.) Now let $V_g:=U_g\otimes
\openone_{\bar n}$ (where $\openone_{\bar n}$ can come either from 
blocking, or from adding extra entangled bonds), and consider the
representation $W'_{\bm g}$, $\bm g\in \tilde G$, generated by
$V_{g_1}\otimes\dots \otimes V_{g_k}$ and the permutation action 
$\Pi_{\bm g}$ on the tensor components.  The character of this
representation is given by
\begin{align*}
\chi_{W'}(\bm g) 
&= \mathrm{tr}[(V_{g_1}\otimes\dots \otimes V_{g_k})\Pi_{\bm g}]
\\
&= \underbrace{
	\mathrm{tr}[(U_{g_1}\otimes\dots \otimes U_{g_k})\Pi_{\bm g}]
    }_{
	=:\chi_W(\bm g)
    }
    \underbrace{
    \mathrm{tr}[(\openone_{\bar n}\otimes\dots \otimes 
			\openone_{\bar n})\Pi_{\bm g}]
    }_{
    =:\chi_\Pi(\bm g)
    }\:.
\end{align*}
We now have that $\chi_W(1)=|G|^k$, and $\chi_W(\bm g)=0$ for all $\bm
g\ne1$
with trivial permutation action, $\Pi_{\bm g}=\Id$; moreover, $|\chi_W(\bm
g)|\le\chi_W(1)=|G|^k$ is independent of $\bar n$. On the other hand,
$\chi_\Pi(\bm g)={\bar n}^c$, where $c$ is the number of cycles in
$\Pi_{\bm g}$, i.e., $\chi_\Pi(\bm g)={\bar n}^k$ for $\Pi_{\bm g}=\Id$,
and $|\chi_\Pi(\bm g)|\le {\bar n}^{k-1}$ otherwise. We thus see that
$\chi_{W'}(1)=|G|^k {\bar n}^k$, while $|\chi_{W'}(\bm g)|\le |G|^k{\bar
n}^{k-1}$ for $\bm g\ne 1$.

Now let $\chi_\alpha(\bm g)$ be the character of an irrep $\alpha$ of
$\tilde G$, with dimension $d_\alpha$. Then, the multiplicity of $\alpha$
in $W'_{\bm g}$ is given by
\begin{align*}
\mu_\alpha &= \frac{1}{|\tilde G|}
    \sum_{\bm g\in\tilde G}\chi^*_{W'}(\bm g)\chi_\alpha(\bm g)
\\
&= \frac{1}{|\tilde G|} \Big[\chi^*_{W'}(1)\chi_\alpha(1) + 
		\sum_{\bm g\ne 1}
    \chi^*_{W'}(\bm g)\chi_\alpha(\bm g)\Big]
\\
&= \frac{1}{|\tilde G|} \Big[|G|^k \bar n^k\, d_\alpha + 
		\sum_{\bm g\ne 1}
    \chi^*_{W'}(\bm g)\chi_\alpha(\bm g)\Big]\ .
\end{align*}
We thus have that 
\[
\left|\mu_\alpha\!-\!\frac{d_\alpha|G|^k{\bar n}^k}{|\tilde G|}\right|
= 
\frac{1}{|\tilde G|}
\Big|\!\sum_{\bm g\ne 1}\chi^*_{W'}(\bm g)\chi_\alpha(\bm g)\Big|
\le 
|G|^k{\bar n}^{k-1} d_\alpha \:,
\]
where we have used that $|\chi_\alpha(\bm g)|\le d_\alpha$.
Thus, in order to obtain the correct relative weights $d_\alpha$ of the
regular representation, the weights $\mu_\alpha$ need to be changed by at
most
\begin{align*}
\rho &= \frac{\mathrm{min}(\mu_\alpha/d_\alpha)}{
		    \mathrm{max}(\mu_\alpha/d_\alpha)}
=\frac{|G|^k{\bar n}^k/|\tilde G|-|G|^k {\bar n}^{k-1}}{
|G|^k{\bar n}^k/|\tilde G|+|G|^k {\bar n}^{k-1}}
\\
&\ge 1 - \frac{2 |\tilde G|}{\bar n}\ .
\end{align*}
As before, this yields that the symmetrized model is in the $D(\tilde G)$
phase if $\rho\ge \rho_0\approx 0.967$, and thus $\bar n\ge 2|\tilde
G|/(1-\rho_0) \approx 60.6\,|G|^k k!$.

\section{Identification of anyon condensation pattern
\label{sec:app:anyons}}

In this appendix, we describe how to identify the different anyon
condensations we find in the symmetrized $D(\mathbb Z_2)$ model.

We begin by setting a notation for the dihedral group $\mathrm{D}_4 = (\mathbb{Z}_2 \times \mathbb{Z}_2) \rtimes \mathbb{Z}_2$, and the anyons of its quantum double.
The symmetry generators of the symmetrized toric code are $X\otimes\Id$, $\Id\otimes X$ (for the two layers), and $\mathbb F$ which flips the layers.
The $\mathrm{D}_4$ group has two generators $x:=\mathbb F$ and $a:=(\Id\otimes X)\mathbb F$
with the group presentation $\langle x,a|x^2=a^4=1,xax^{-1}=a^{-1}\rangle$.
The eight group elements partition into the five conjugacy classes
\begin{align}\begin{split}
C_1 &= \{e\}\,,\\
C_2 &= \{a^2\}\,,\\
C_3 &= \{a,a^3\}\,,\\
C_4 &= \{x,a^2x\}\,,\\
C_5 &= \{ax,a^3x\}\,.\\
\end{split}\end{align}
Let us now consider the irreps of their normalizers $N[C]$ (defined as
equivalence classes of the isomorphic normalizers $N[g]=\{h\in G:gh=hg\}$,
$g\in C$) which together with $C_n$ label the particle sectors:
\begin{itemize}
\item
$N[C_1]=N[C_2]=\mathrm{D}_4$, with irrep characters
\[
	\begin{tabular}{l|c|c|c|c|c|}
	    &  $C_1$ & $C_2$ & $C_3$ & $C_4$ & $C_5$ \\
	\hline
	$\alpha_0$ &  $1$ &  $1$&   $1$&   $1$&  $1$\\
	\hline
	$\alpha_1$ &  $1$ &  $1$&   $1$&  $-1$& $-1$\\
	\hline
	$\alpha_2$ &  $1$ &  $1$&  $-1$&   $1$& $-1$\\
	\hline
	$\alpha_3$ &  $1$ &  $1$&  $-1$&  $-1$&  $1$\\
	\hline
	$\alpha_4$ &  $2$ & $-2$&   $0$&   $0$&  $0$\\
	\hline
	\end{tabular}
\]
\item 
$N[C_3]\cong N[a]=\{e,a,a^2,a^3\}\cong \mathbb Z_4$
with irreps $\alpha_k(a) = i^{k}$.
\item 
$N[C_4]\cong N[x]=\{e,x,a^2,a^2x\}\cong \mathbb Z_2\times \mathbb Z_2$
with irreps $\alpha_k(x) = (-1)^{k_1}$,
$\alpha_k(a^2) = (-1)^{k_2}$, $k=2k_2+k_1$.
\item 
$N[C_5]\cong N[ax]=\{e,ax,a^2,a^3x\}\cong \mathbb Z_2\times \mathbb Z_2$
with irreps $\alpha_k(ax) = (-1)^{k_1}$,
$\alpha_k(a^2) = (-1)^{k_2}$, $k=2k_2+k_1$.
\end{itemize}
The irreps are given for the correspondingly listed element of the
conjugacy class.  Each anyon of $D(\mathrm{D}_4)$ is labelled by one
conjugacy class $C_n$ and an irrep $\alpha_m$ of its normalizer.  We will
employ the shorthand notation $nm$, where $n=1,\dots,5$, and
$m=a,b,c,\dots$ ($a\equiv0$, $b\equiv1$, etc.) to label the 22 anyons.
For example, $4b$ labels the anyon with conjugacy class $C_4$ and irrep
$\alpha_1$; in particular, $1a$ is the trivial (vacuum) anyon.
Table~\ref{tab:D4_anyons} lists the anyons.

\begin{table*}[t]
	\begin{align*}
	\begin{array}{ c | *{22}{c} }
	      \text{Label}       & 1a & 1b & 1c & 1d & 1e & 2a & 2b & 2c & 2d & 2e & 3a & 3b & 3c & 3d & 4a & 4b & 4c & 4d & 5a & 5b & 5c & 5d \\
	\hline
	\text{Quantum dimension} & 1  & 1  & 1  & 1  & 2  & 1  & 1  & 1  & 1  & 2  & 2  & 2  & 2  & 2  & 2  & 2  & 2  & 2  & 2  & 2  & 2  & 2 \\
	\text{Topological spin}  & 1  & 1  & 1  & 1  & 1  & 1  & 1  & 1  & 1  & -1 & 1  & i  & -1 & -i & 1  & -1 & 1  & -1 & 1  & -1 & 1  & -1
	\end{array}
	\end{align*}
	\caption{\label{tab:D4_anyons}
The anyons of $D(\mathrm{D}_4)$ and their properties.  Each anyon is its
own anti-particle.  The topological spin is given by $\alpha_m(g)$, where
$g$ is the representative of the conjugacy class $C_n$ used to define the
irrep characters $\alpha_m$ of $N[g]\cong N[C]$.  
}
\end{table*}

On an infinitely-long cylinder (or torus), there is a one-to-one correspondence
between anyons and ground states; we denote these states as $\ket{\psi_A}$
where $A$ is the anyon label.  Physically, these states are constructed by
starting with the wavefunction corresponding to the vacuum
$\ket{\psi_{1a}}$, creating a pair of $A$-$A$ anyons and pulling them
apart to opposite ends of the cylinder.

In order to identify the anyon condensation pattern, we compute the
overlap between all possible states $\ket{\psi_A}$ in the thermodynamic
limit.  Several things can
happen~\cite{schuch:topo-top,haegeman:shadows,duivenvoorden:anyon-condensation}.
If \emph{all} the states have non-vanishing norm and remain orthogonal to
one another, this implies that no anyon condensation occured and we remain
in the $D(\mathrm{D}_4)$ topological phase.  However, if the norm of any
state $\ket{\psi_A}$ does go to zero (relative to that of
$\ket{\psi_{1a}}$), then this corresponds to the case where the anyon $A$
has becomes confined, indicative of an anyon condensation process.
Another possibility is that the overlap of two states $\ket{\psi_A}$ and
$\ket{\psi_B}$ goes to a non-zero constant rather then being orthogonal;
in such case, the anyons of the condensed theory are constructed from
superpositions of $A$ and $B$.  In particular, an anyon $C$ is condensed
if it forms a superposition with the vacuum.

\begin{figure}[b]
\includegraphics[scale=1]{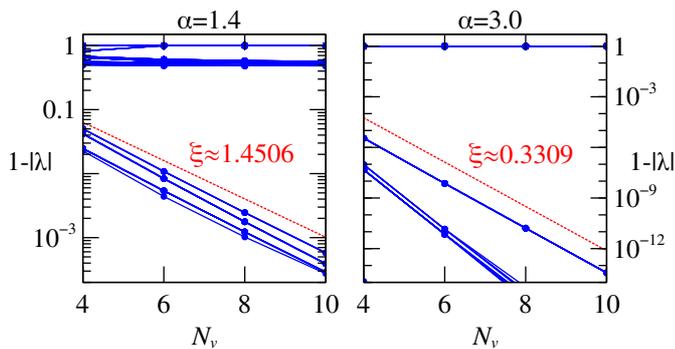}
\caption{\label{fig:sectorscaling}
Overlap per column $\lambda_{N_v}\equiv \lambda$ (on a cylinder of
circumference $N_v$) of anyon sectors for the $\alpha$-interpolation
\eqref{eq:alpha-interpol-p5} of SYM$_{\nicefrac12\times\nicefrac12}$. One
can clearly distinguish the $\lambda$ which converge to $1$ exponentially
from those which go to a constant. For comparison, we plot $c\exp(-N_v/\xi)$
with $\xi$ the largest correlation length of Fig.~\ref{fig:p5xp5-alpha}a,
which confirms that the overlap of ground state sectors is govered by the
anyon-anyon correlations.  Since there are $253$ different lines, we
omit labels; the condensed anyon theory extracted from the data is given
in the text.} 
\vspace*{-1em}
\end{figure}

In order to numerically compute the overlap, we put $\ket{\psi_A}$ on a
long cylinder (or torus) of length $N_h$ and circumference $N_v$.  There,
a state $\ket{\psi_A}$ with $A\equiv nm$ is constructed by (i) placing a
string of $U_g$ for some $g\in C_n$ along the cylinder axis when closing
the boundary, and (ii) by projecting the boundary of the cylinder onto the
irrep sector $\alpha_m$ of $N[g]\cong N[C_n]$;
cf.~Refs.~\cite{schuch:peps-sym,schuch:topo-top} for details. The overlap
of two states is then the overlap of the corresponding PEPS, which 
asymptotically scales like the $N_h$\textsuperscript{th} power of the
largest eigenvalue $\lambda_{N_v}$ of one column of the corresponding
mixed transfer operator, cf.~Ref.~\cite{schuch:topo-top}.  We therefore
need to analyze whether $\lambda_{N_v}$ (normalized by the largest
eigenvalue in the trivial sector) converges to one: If it does (at a
sufficient rate), then $\lambda_{N_v}^{N_h}\rightarrow 1$ for a coupled
limit $N_h,N_v\rightarrow\infty$, while if it doesn't,
$\lambda_{N_v}^{N_h}\rightarrow 0$. To this end, we use exact
diagonalization of the transfer operator together with finite size
scaling, cf.~Ref.~\cite{schuch:topo-top}.

Figure~\ref{fig:sectorscaling} shows the corresponding data for the two
phases at $\alpha=1.4$ and $\alpha=3.0$ in the $\alpha$-interpolation
[Eq.~\eqref{eq:alpha-interpol-p5}] of the 
SYM$_{\nicefrac12\times\nicefrac12}$ model, Fig.~\ref{fig:p5xp5-alpha}a,
and we find that we can clearly distinguish the two different scaling
behaviors of $\lambda_{N_v}$. For the small $\alpha$ phase, we consider $\alpha=0$,
which is a fixed point wavefunction and thus $\lambda_{N_v}\in\{0,1\}$
independent of $N_v$. We can now study the condensation and confinement of
anyons as described above.  For $\alpha=0$, we find that the following
ground states have non-trivial overlaps within each set:
$(1a,1c,4a)$, $(2a,2c,4a)$, $(1e,4c)$, $(2e,4d)$, while all other
wavefunctions have vanishing norm~%
\footnote{Note that the largest eigenvalue of the transfer operator for the
normalization {$\langle\psi_{4a}\vert\psi_{4a}\rangle$} of the non-abelian
anyon $4a$ is degenerate, which allows it to have non-zero overlap with
more than one sector (while without such a degeneracy, i.e., for abelian
anyons, the normalized overlap must always be zero or one).
}.
From this we have the anyon condensation process:
\[
\begin{array}{l@{\quad\rightarrow\quad}l}
1a + 1c + 4a & \hat 1\\
1e + 4c	    & \hat e\\
2a + 2c + 4a & \hat m \\
2e + 4d	    & \hat f \ .
\end{array}
\]
Indeed, the resulting phase has toric code topological order: The anyons
$\hat{1}$, $\hat{e}$, and $\hat{m}$ are bosons (i.e., their components
have topological spin $+1$), while $\hat{f}$ is a fermion (with
topological spin $-1$).  The mutual statistics of $\hat{e}$, $\hat{m}$,
and $\hat{f}$ are fermionic (i.e., their mutual $S$-matrix elements are
$-1$,
cf.~Refs.~\cite{bais:anyon-condensation,neupert:boson-condensation}).  We
can also understand this condensation as a two step process:
$D(\mathrm{D}_4) \rightarrow D(\mathbb{Z}_2\times \mathbb Z_2) \rightarrow
D(\mathbb{Z}_2)$.  Condensing $1c$ generates the toric code squared
$D(\mathbb Z_2\times \mathbb Z_2)$, which splits the $4a$ anyon into two
abelian anyons, the second step condenses one of the split $4a$ anyon to
yield the toric code order.  As $\alpha$ increases, the correlation length
also increases (Fig.~\ref{fig:p5xp5-alpha}) but the topological order
remain the same up to the phase transition.  

For $\alpha=1.4$, we find the condensation pattern
\[
\vspace*{-1em}
\begin{array}{l@{\quad\rightarrow\quad}l}
1a + 2a + 4a & \hat 1\\
1c + 2c + 4a & \hat e \\
3a + 5a	    & \hat m  \\
3c + 5b      & \hat f\ .
\end{array}
\vspace*{-1em}
\]
Again, this phase has toric code topological order, but with a different condensation pattern, and thus inequivalent to the toric code found in the small $\alpha$ phase.
Consequently, a phase transition must occur at some intermediate $\alpha$ (which we found numerically at $\alpha\approx1$) if we are to preserve the $\mathrm{D}_4$ symmetry of the PEPS.
Note that the two condensation patterns above are exchanged if we choose
to apply $\Pi(\alpha)$, Eq.~\eqref{eq:alpha-interpol-p5}, in the dual basis.
Finally, for $\alpha=3$, we find that anyons $1a,1c,2a,2c,4a$ have
condensed into the vacuum \footnote{The exact condensation pattern is given by $1a+1c+2a+2c+4a+4a \rightarrow\hat{1}$, with $4a$ having twice the weight as compared to the other anyons.}, while all other anyons have become confined; we
are thus left with a trivial phase, as expected.

We have applied the same analysis also to the $\alpha$-interpolation for
the SYM$_{1\times 1}$ model described in Eq.~\eqref{eq:alpha-interpol-1}.
For the small $\alpha$ phase, we can again consider $\alpha=0$ and find a toric
code with the same condensation pattern as before for $\alpha=1.4$ (since
we chose to symmetrize in the dual basis). For the large $\alpha$ regime,
however, we cannot reliably extract the scaling of the $\lambda_{N_v}$ due to
the large correlation length.  In order to identify the condensation
pattern, we therefore choose an alternative approach: We consider an
infinite plane and compute the overlap of different anyons (with a
semi-infinite string attached to them), using an iMPS ansatz,
cf.~Ref.~\cite{haegeman:shadows}.  Here, a non-abelian anyon is described
by a semi-infinite string of $U_g$ with $g\in C_n$, terminated by an object
$R_{\alpha_m}$ which transforms like an irrep $\alpha_m$ of $N[g]$; in
order to account for the symmetry breaking in the boundary
MPS~\cite{haegeman:shadows,duivenvoorden:anyon-condensation}, we
additionally need to symmetrize the iMPS over the group action, since the
group is non-abelian. In order to not have to worry about the precise
choice for $R_\alpha$, and in order to reduce computational effort, we
choose to rather compute the boundary conditions imposed on an arbitrary
endpoint $R_\alpha\otimes \bar R_\beta$ in the overlap, and subsequently
project it onto all possible irrep sectors $(\alpha,\beta)$; details of
the method will be presented elsewhere. This way, we find that all anyons
describe well-normalized and orthogonal excitations, suggesting that the
large $\alpha$ regime corresponds to the full $D(\mathrm{D}_4)$ phase.


\vspace*{-0.4cm}
\begin{thebibliography}{31}%
\makeatletter
\providecommand \@ifxundefined [1]{%
 \@ifx{#1\undefined}
}%
\providecommand \@ifnum [1]{%
 \ifnum #1\expandafter \@firstoftwo
 \else \expandafter \@secondoftwo
 \fi
}%
\providecommand \@ifx [1]{%
 \ifx #1\expandafter \@firstoftwo
 \else \expandafter \@secondoftwo
 \fi
}%
\providecommand \natexlab [1]{#1}%
\providecommand \enquote  [1]{``#1''}%
\providecommand \bibnamefont  [1]{#1}%
\providecommand \bibfnamefont [1]{#1}%
\providecommand \citenamefont [1]{#1}%
\providecommand \href@noop [0]{\@secondoftwo}%
\providecommand \href [0]{\begingroup \@sanitize@url \@href}%
\providecommand \@href[1]{\@@startlink{#1}\@@href}%
\providecommand \@@href[1]{\endgroup#1\@@endlink}%
\providecommand \@sanitize@url [0]{\catcode `\\12\catcode `\$12\catcode
  `\&12\catcode `\#12\catcode `\^12\catcode `\_12\catcode `\%12\relax}%
\providecommand \@@startlink[1]{}%
\providecommand \@@endlink[0]{}%
\providecommand \url  [0]{\begingroup\@sanitize@url \@url }%
\providecommand \@url [1]{\endgroup\@href {#1}{\urlprefix }}%
\providecommand \urlprefix  [0]{URL }%
\providecommand \Eprint [0]{\href }%
\providecommand \doibase [0]{http://dx.doi.org/}%
\providecommand \selectlanguage [0]{\@gobble}%
\providecommand \bibinfo  [0]{\@secondoftwo}%
\providecommand \bibfield  [0]{\@secondoftwo}%
\providecommand \translation [1]{[#1]}%
\providecommand \BibitemOpen [0]{}%
\providecommand \bibitemStop [0]{}%
\providecommand \bibitemNoStop [0]{.\EOS\space}%
\providecommand \EOS [0]{\spacefactor3000\relax}%
\providecommand \BibitemShut  [1]{\csname bibitem#1\endcsname}%
\let\auto@bib@innerbib\@empty
\bibitem [{\citenamefont {Bais}\ and\ \citenamefont
  {Slingerland}(2009)}]{bais:anyon-condensation}%
  \BibitemOpen
  \bibfield  {author} {\bibinfo {author} {\bibfnamefont {F.}~\bibnamefont
  {Bais}}\ and\ \bibinfo {author} {\bibfnamefont {J.}~\bibnamefont
  {Slingerland}},\ }\href@noop {} {\bibfield  {journal} {\bibinfo  {journal}
  {Phys. Rev. B}\ }\textbf {\bibinfo {volume} {79}},\ \bibinfo {pages} {045316}
  (\bibinfo {year} {2009})},\ \Eprint {http://arxiv.org/abs/arXiv:0808.0627}
  {arXiv:0808.0627} \BibitemShut {NoStop}%
\bibitem [{\citenamefont {{Eli{\"e}ns}}\ \emph {et~al.}(2014)\citenamefont
  {{Eli{\"e}ns}}, \citenamefont {{Romers}},\ and\ \citenamefont
  {{Bais}}}]{eliens:diagrammatic-anyon-condensation}%
  \BibitemOpen
  \bibfield  {author} {\bibinfo {author} {\bibfnamefont {I.~S.}\ \bibnamefont
  {{Eli{\"e}ns}}}, \bibinfo {author} {\bibfnamefont {J.~C.}\ \bibnamefont
  {{Romers}}}, \ and\ \bibinfo {author} {\bibfnamefont {F.~A.}\ \bibnamefont
  {{Bais}}},\ }\href {\doibase 10.1103/PhysRevB.90.195130} {\bibfield
  {journal} {\bibinfo  {journal} {\prb}\ }\textbf {\bibinfo {volume} {90}},\
  \bibinfo {eid} {195130} (\bibinfo {year} {2014})},\ \Eprint
  {http://arxiv.org/abs/1310.6001} {arXiv:1310.6001} \BibitemShut {NoStop}%
\bibitem [{\citenamefont {{Neupert}}\ \emph {et~al.}(2016)\citenamefont
  {{Neupert}}, \citenamefont {{He}}, \citenamefont {{von Keyserlingk}},
  \citenamefont {{Sierra}},\ and\ \citenamefont
  {{Bernevig}}}]{neupert:boson-condensation}%
  \BibitemOpen
  \bibfield  {author} {\bibinfo {author} {\bibfnamefont {T.}~\bibnamefont
  {{Neupert}}}, \bibinfo {author} {\bibfnamefont {H.}~\bibnamefont {{He}}},
  \bibinfo {author} {\bibfnamefont {C.}~\bibnamefont {{von Keyserlingk}}},
  \bibinfo {author} {\bibfnamefont {G.}~\bibnamefont {{Sierra}}}, \ and\
  \bibinfo {author} {\bibfnamefont {B.~A.}\ \bibnamefont {{Bernevig}}},\ }\href
  {\doibase 10.1103/PhysRevB.93.115103} {\bibfield  {journal} {\bibinfo
  {journal} {\prb}\ }\textbf {\bibinfo {volume} {93}},\ \bibinfo {eid} {115103}
  (\bibinfo {year} {2016})},\ \Eprint {http://arxiv.org/abs/1601.01320}
  {arXiv:1601.01320} \BibitemShut {NoStop}%
\bibitem [{\citenamefont {{Read}}\ and\ \citenamefont
  {{Rezayi}}(1999)}]{read:projective}%
  \BibitemOpen
  \bibfield  {author} {\bibinfo {author} {\bibfnamefont {N.}~\bibnamefont
  {{Read}}}\ and\ \bibinfo {author} {\bibfnamefont {E.}~\bibnamefont
  {{Rezayi}}},\ }\href {\doibase 10.1103/PhysRevB.59.8084} {\bibfield
  {journal} {\bibinfo  {journal} {\prb}\ }\textbf {\bibinfo {volume} {59}},\
  \bibinfo {pages} {8084} (\bibinfo {year} {1999})},\ \Eprint
  {http://arxiv.org/abs/cond-mat/9809384} {cond-mat/9809384} \BibitemShut
  {NoStop}%
\bibitem [{\citenamefont {{Cappelli}}\ \emph {et~al.}(2001)\citenamefont
  {{Cappelli}}, \citenamefont {{Georgiev}},\ and\ \citenamefont
  {{Todorov}}}]{cappelli:coset}%
  \BibitemOpen
  \bibfield  {author} {\bibinfo {author} {\bibfnamefont {A.}~\bibnamefont
  {{Cappelli}}}, \bibinfo {author} {\bibfnamefont {L.~S.}\ \bibnamefont
  {{Georgiev}}}, \ and\ \bibinfo {author} {\bibfnamefont {I.~T.}\ \bibnamefont
  {{Todorov}}},\ }\href {\doibase 10.1016/S0550-3213(00)00774-4} {\bibfield
  {journal} {\bibinfo  {journal} {Nuclear Physics B}\ }\textbf {\bibinfo
  {volume} {599}},\ \bibinfo {pages} {499} (\bibinfo {year} {2001})},\ \Eprint
  {http://arxiv.org/abs/hep-th/0009229} {hep-th/0009229} \BibitemShut {NoStop}%
\bibitem [{\citenamefont {Kitaev}(2003)}]{kitaev:toriccode}%
  \BibitemOpen
  \bibfield  {author} {\bibinfo {author} {\bibfnamefont {A.}~\bibnamefont
  {Kitaev}},\ }\href@noop {} {\bibfield  {journal} {\bibinfo  {journal} {Ann.
  Phys.}\ }\textbf {\bibinfo {volume} {303}},\ \bibinfo {pages} {2} (\bibinfo
  {year} {2003})},\ \Eprint {http://arxiv.org/abs/quant-ph/9707021}
  {quant-ph/9707021} \BibitemShut {NoStop}%
\bibitem [{\citenamefont {Paredes}(2012)}]{paredes:sym-tcode-1}%
  \BibitemOpen
  \bibfield  {author} {\bibinfo {author} {\bibfnamefont {B.}~\bibnamefont
  {Paredes}},\ }\href {\doibase 10.1103/PhysRevB.86.155122} {\bibfield
  {journal} {\bibinfo  {journal} {Phys. Rev. B}\ }\textbf {\bibinfo {volume}
  {86}},\ \bibinfo {pages} {155122} (\bibinfo {year} {2012})},\ \Eprint
  {http://arxiv.org/abs/arXiv:1209.6040} {arXiv:1209.6040} \BibitemShut
  {NoStop}%
\bibitem [{\citenamefont {Paredes}(2014)}]{paredes:sym-tcode-2}%
  \BibitemOpen
  \bibfield  {author} {\bibinfo {author} {\bibfnamefont {B.}~\bibnamefont
  {Paredes}},\ }\href@noop {} {\  (\bibinfo {year} {2014})},\ \Eprint
  {http://arxiv.org/abs/arXiv:1402.3567} {arXiv:1402.3567} \BibitemShut
  {NoStop}%
\bibitem [{\citenamefont {Verstraete}\ and\ \citenamefont
  {Cirac}(2004)}]{verstraete:mbc-peps}%
  \BibitemOpen
  \bibfield  {author} {\bibinfo {author} {\bibfnamefont {F.}~\bibnamefont
  {Verstraete}}\ and\ \bibinfo {author} {\bibfnamefont {J.~I.}\ \bibnamefont
  {Cirac}},\ }\href@noop {} {\bibfield  {journal} {\bibinfo  {journal}
  {Phys.~Rev.~A}\ }\textbf {\bibinfo {volume} {70}},\ \bibinfo {pages} {060302}
  (\bibinfo {year} {2004})},\ \Eprint {http://arxiv.org/abs/quant-ph/0311130}
  {quant-ph/0311130} \BibitemShut {NoStop}%
\bibitem [{\citenamefont {Verstraete}\ \emph {et~al.}(2006)\citenamefont
  {Verstraete}, \citenamefont {Wolf}, \citenamefont {Perez-Garcia},\ and\
  \citenamefont {Cirac}}]{verstraete:comp-power-of-peps}%
  \BibitemOpen
  \bibfield  {author} {\bibinfo {author} {\bibfnamefont {F.}~\bibnamefont
  {Verstraete}}, \bibinfo {author} {\bibfnamefont {M.~M.}\ \bibnamefont
  {Wolf}}, \bibinfo {author} {\bibfnamefont {D.}~\bibnamefont {Perez-Garcia}},
  \ and\ \bibinfo {author} {\bibfnamefont {J.~I.}\ \bibnamefont {Cirac}},\
  }\href@noop {} {\bibfield  {journal} {\bibinfo  {journal} {Phys.\ Rev.\
  Lett.}\ }\textbf {\bibinfo {volume} {96}},\ \bibinfo {pages} {220601}
  (\bibinfo {year} {2006})},\ \Eprint {http://arxiv.org/abs/quant-ph/0601075}
  {quant-ph/0601075} \BibitemShut {NoStop}%
\bibitem [{\citenamefont {{Schuch}}\ \emph {et~al.}(2010)\citenamefont
  {{Schuch}}, \citenamefont {{Cirac}},\ and\ \citenamefont
  {{P{\'e}rez-Garc{\'{\i}}a}}}]{schuch:peps-sym}%
  \BibitemOpen
  \bibfield  {author} {\bibinfo {author} {\bibfnamefont {N.}~\bibnamefont
  {{Schuch}}}, \bibinfo {author} {\bibfnamefont {I.}~\bibnamefont {{Cirac}}}, \
  and\ \bibinfo {author} {\bibfnamefont {D.}~\bibnamefont
  {{P{\'e}rez-Garc{\'{\i}}a}}},\ }\href {\doibase 10.1016/j.aop.2010.05.008}
  {\bibfield  {journal} {\bibinfo  {journal} {Ann. Phys.}\ }\textbf {\bibinfo
  {volume} {325}},\ \bibinfo {pages} {2153} (\bibinfo {year} {2010})},\ \Eprint
  {http://arxiv.org/abs/arXiv:1001.3807} {arXiv:1001.3807} \BibitemShut
  {NoStop}%
\bibitem [{\citenamefont {Levin}\ and\ \citenamefont
  {Wen}(2005)}]{levin:stringnets}%
  \BibitemOpen
  \bibfield  {author} {\bibinfo {author} {\bibfnamefont {M.~A.}\ \bibnamefont
  {Levin}}\ and\ \bibinfo {author} {\bibfnamefont {X.-G.}\ \bibnamefont
  {Wen}},\ }\href@noop {} {\bibfield  {journal} {\bibinfo  {journal} {Phys.
  Rev. B}\ }\textbf {\bibinfo {volume} {71}},\ \bibinfo {pages} {045110}
  (\bibinfo {year} {2005})},\ \Eprint {http://arxiv.org/abs/cond-mat/0404617}
  {cond-mat/0404617} \BibitemShut {NoStop}%
\bibitem [{\citenamefont {Buerschaper}\ \emph {et~al.}(2009)\citenamefont
  {Buerschaper}, \citenamefont {Aguado},\ and\ \citenamefont
  {Vidal}}]{buerschaper:stringnet-peps}%
  \BibitemOpen
  \bibfield  {author} {\bibinfo {author} {\bibfnamefont {O.}~\bibnamefont
  {Buerschaper}}, \bibinfo {author} {\bibfnamefont {M.}~\bibnamefont {Aguado}},
  \ and\ \bibinfo {author} {\bibfnamefont {G.}~\bibnamefont {Vidal}},\
  }\href@noop {} {\bibfield  {journal} {\bibinfo  {journal} {Phys.\ Rev. B}\
  }\textbf {\bibinfo {volume} {79}},\ \bibinfo {pages} {085119} (\bibinfo
  {year} {2009})},\ \Eprint {http://arxiv.org/abs/arXiv:0809.2393}
  {arXiv:0809.2393} \BibitemShut {NoStop}%
\bibitem [{\citenamefont {Gu}\ \emph {et~al.}(2009)\citenamefont {Gu},
  \citenamefont {Levin}, \citenamefont {Swingle},\ and\ \citenamefont
  {Wen}}]{gu:stringnet-peps}%
  \BibitemOpen
  \bibfield  {author} {\bibinfo {author} {\bibfnamefont {Z.-C.}\ \bibnamefont
  {Gu}}, \bibinfo {author} {\bibfnamefont {M.}~\bibnamefont {Levin}}, \bibinfo
  {author} {\bibfnamefont {B.}~\bibnamefont {Swingle}}, \ and\ \bibinfo
  {author} {\bibfnamefont {X.-G.}\ \bibnamefont {Wen}},\ }\href@noop {}
  {\bibfield  {journal} {\bibinfo  {journal} {Phys. Rev. B}\ }\textbf {\bibinfo
  {volume} {79}},\ \bibinfo {pages} {085118} (\bibinfo {year} {2009})},\
  \Eprint {http://arxiv.org/abs/arXiv:0809.2821} {arXiv:0809.2821} \BibitemShut
  {NoStop}%
\bibitem [{\citenamefont {Sahinoglu}\ \emph {et~al.}(2014)\citenamefont
  {Sahinoglu}, \citenamefont {Williamson}, \citenamefont {Bultinck},
  \citenamefont {Marien}, \citenamefont {Haegeman}, \citenamefont {Schuch},\
  and\ \citenamefont {Verstraete}}]{sahinoglu:mpo-injectivity}%
  \BibitemOpen
  \bibfield  {author} {\bibinfo {author} {\bibfnamefont {M.~B.}\ \bibnamefont
  {Sahinoglu}}, \bibinfo {author} {\bibfnamefont {D.}~\bibnamefont
  {Williamson}}, \bibinfo {author} {\bibfnamefont {N.}~\bibnamefont
  {Bultinck}}, \bibinfo {author} {\bibfnamefont {M.}~\bibnamefont {Marien}},
  \bibinfo {author} {\bibfnamefont {J.}~\bibnamefont {Haegeman}}, \bibinfo
  {author} {\bibfnamefont {N.}~\bibnamefont {Schuch}}, \ and\ \bibinfo {author}
  {\bibfnamefont {F.}~\bibnamefont {Verstraete}},\ }\href@noop {} {\  (\bibinfo
  {year} {2014})},\ \Eprint {http://arxiv.org/abs/arXiv:1409.2150}
  {arXiv:1409.2150} \BibitemShut {NoStop}%
\bibitem [{\citenamefont {{Bultinck}}\ \emph {et~al.}(2015)\citenamefont
  {{Bultinck}}, \citenamefont {{Mari{\"e}n}}, \citenamefont {{Williamson}},
  \citenamefont {{{\c S}ahino{\u g}lu}}, \citenamefont {{Haegeman}},\ and\
  \citenamefont {{Verstraete}}}]{bultinck:mpo-anyons}%
  \BibitemOpen
  \bibfield  {author} {\bibinfo {author} {\bibfnamefont {N.}~\bibnamefont
  {{Bultinck}}}, \bibinfo {author} {\bibfnamefont {M.}~\bibnamefont
  {{Mari{\"e}n}}}, \bibinfo {author} {\bibfnamefont {D.~J.}\ \bibnamefont
  {{Williamson}}}, \bibinfo {author} {\bibfnamefont {M.~B.}\ \bibnamefont {{{\c
  S}ahino{\u g}lu}}}, \bibinfo {author} {\bibfnamefont {J.}~\bibnamefont
  {{Haegeman}}}, \ and\ \bibinfo {author} {\bibfnamefont {F.}~\bibnamefont
  {{Verstraete}}},\ }\href@noop {} {\  (\bibinfo {year} {2015})},\ \Eprint
  {http://arxiv.org/abs/arXiv:1511.08090} {arXiv:1511.08090} \BibitemShut
  {NoStop}%
\bibitem [{\citenamefont {Schuch}\ \emph {et~al.}(2013)\citenamefont {Schuch},
  \citenamefont {Poilblanc}, \citenamefont {Cirac},\ and\ \citenamefont
  {Perez-Garcia}}]{schuch:topo-top}%
  \BibitemOpen
  \bibfield  {author} {\bibinfo {author} {\bibfnamefont {N.}~\bibnamefont
  {Schuch}}, \bibinfo {author} {\bibfnamefont {D.}~\bibnamefont {Poilblanc}},
  \bibinfo {author} {\bibfnamefont {J.~I.}\ \bibnamefont {Cirac}}, \ and\
  \bibinfo {author} {\bibfnamefont {D.}~\bibnamefont {Perez-Garcia}},\
  }\href@noop {} {\bibfield  {journal} {\bibinfo  {journal} {Phys. Rev. Lett.}\
  }\textbf {\bibinfo {volume} {111}},\ \bibinfo {pages} {090501} (\bibinfo
  {year} {2013})},\ \Eprint {http://arxiv.org/abs/arXiv:1210.5601}
  {arXiv:1210.5601} \BibitemShut {NoStop}%
\bibitem [{\citenamefont {Haegeman}\ \emph {et~al.}(2015)\citenamefont
  {Haegeman}, \citenamefont {Zauner}, \citenamefont {Schuch},\ and\
  \citenamefont {Verstraete}}]{haegeman:shadows}%
  \BibitemOpen
  \bibfield  {author} {\bibinfo {author} {\bibfnamefont {J.}~\bibnamefont
  {Haegeman}}, \bibinfo {author} {\bibfnamefont {V.}~\bibnamefont {Zauner}},
  \bibinfo {author} {\bibfnamefont {N.}~\bibnamefont {Schuch}}, \ and\ \bibinfo
  {author} {\bibfnamefont {F.}~\bibnamefont {Verstraete}},\ }\href {\doibase
  doi:10.1038/ncomms9284} {\bibfield  {journal} {\bibinfo  {journal} {Nature
  Comm.}\ }\textbf {\bibinfo {volume} {6}},\ \bibinfo {pages} {8284} (\bibinfo
  {year} {2015})},\ \Eprint {http://arxiv.org/abs/arXiv:1410.5443}
  {arXiv:1410.5443} \BibitemShut {NoStop}%
\bibitem [{\citenamefont {Duivenvoorden}\ \emph {et~al.}()\citenamefont
  {Duivenvoorden}, \citenamefont {Iqbal}, \citenamefont {Haegeman},
  \citenamefont {Verstraete},\ and\ \citenamefont
  {Schuch}}]{duivenvoorden:anyon-condensation}%
  \BibitemOpen
  \bibfield  {author} {\bibinfo {author} {\bibfnamefont {K.}~\bibnamefont
  {Duivenvoorden}}, \bibinfo {author} {\bibfnamefont {M.}~\bibnamefont
  {Iqbal}}, \bibinfo {author} {\bibfnamefont {J.}~\bibnamefont {Haegeman}},
  \bibinfo {author} {\bibfnamefont {F.}~\bibnamefont {Verstraete}}, \ and\
  \bibinfo {author} {\bibfnamefont {N.}~\bibnamefont {Schuch}},\ }\href@noop {}
  {\bibinfo  {journal} {in preparation}\ }\BibitemShut {NoStop}%
\bibitem [{\citenamefont {Schuch}\ \emph {et~al.}(2012)\citenamefont {Schuch},
  \citenamefont {Poilblanc}, \citenamefont {Cirac},\ and\ \citenamefont
  {P{\'e}rez-Garc{\'\i}a}}]{schuch:rvb-kagome}%
  \BibitemOpen
\bibfield  {journal} {  }\bibfield  {author} {\bibinfo {author} {\bibfnamefont
  {N.}~\bibnamefont {Schuch}}, \bibinfo {author} {\bibfnamefont
  {D.}~\bibnamefont {Poilblanc}}, \bibinfo {author} {\bibfnamefont {J.~I.}\
  \bibnamefont {Cirac}}, \ and\ \bibinfo {author} {\bibfnamefont
  {D.}~\bibnamefont {P{\'e}rez-Garc{\'\i}a}},\ }\href@noop {} {\bibfield
  {journal} {\bibinfo  {journal} {Phys. Rev. B}\ }\textbf {\bibinfo {volume}
  {86}},\ \bibinfo {pages} {115108} (\bibinfo {year} {2012})},\ \Eprint
  {http://arxiv.org/abs/arXiv:1203.4816} {arXiv:1203.4816} \BibitemShut
  {NoStop}%
\bibitem [{\citenamefont {{Schuch}}\ \emph {et~al.}(2011)\citenamefont
  {{Schuch}}, \citenamefont {{Perez-Garcia}},\ and\ \citenamefont
  {{Cirac}}}]{schuch:mps-phases}%
  \BibitemOpen
  \bibfield  {author} {\bibinfo {author} {\bibfnamefont {N.}~\bibnamefont
  {{Schuch}}}, \bibinfo {author} {\bibfnamefont {D.}~\bibnamefont
  {{Perez-Garcia}}}, \ and\ \bibinfo {author} {\bibfnamefont {I.}~\bibnamefont
  {{Cirac}}},\ }\href@noop {} {\bibfield  {journal} {\bibinfo  {journal} {Phys.
  Rev. B}\ }\textbf {\bibinfo {volume} {84}},\ \bibinfo {pages} {165139}
  (\bibinfo {year} {2011})},\ \Eprint {http://arxiv.org/abs/arXiv:1010.3732}
  {arXiv:1010.3732} \BibitemShut {NoStop}%
\bibitem [{Note1()}]{Note1}%
  \BibitemOpen
  \bibinfo {note} {Note that the converse is not true: There can be tensors of
  the form Eq.~\protect \textup {\hbox {\mathsurround \z@ \protect \normalfont
  (\ignorespaces \ref {eq:def-Giso-tensor}\unskip \@@italiccorr )}} where the
  irrep weights in $U_g$ are very far from the regular representation, but
  which nevertheless are in the $D(G)$ phase; this is related to the fact that
  missing irreps can be obtained by blocking (i.e., as irreps of $U_g\otimes
  U_g$). For instance, this is the case for the model obtained from the
  $D(\protect \mathrm {D}_4)$ model by removing the trivial irrep from the
  regular representation.}\BibitemShut {Stop}%
\bibitem [{\citenamefont {Repellin}\ \emph {et~al.}(2015)\citenamefont
  {Repellin}, \citenamefont {Neupert}, \citenamefont {Bernevig},\ and\
  \citenamefont {Regnault}}]{repellin:Zk-groundstates}%
  \BibitemOpen
  \bibfield  {author} {\bibinfo {author} {\bibfnamefont {C.}~\bibnamefont
  {Repellin}}, \bibinfo {author} {\bibfnamefont {T.}~\bibnamefont {Neupert}},
  \bibinfo {author} {\bibfnamefont {B.~A.}\ \bibnamefont {Bernevig}}, \ and\
  \bibinfo {author} {\bibfnamefont {N.}~\bibnamefont {Regnault}},\ }\href@noop
  {} {\bibfield  {journal} {\bibinfo  {journal} {Phys. Rev. B}\ }\textbf
  {\bibinfo {volume} {92}},\ \bibinfo {pages} {115128} (\bibinfo {year}
  {2015})},\ \Eprint {http://arxiv.org/abs/arXiv:1504.00367} {arXiv:1504.00367}
  \BibitemShut {NoStop}%
\bibitem [{\citenamefont {Fern\'andez-Gonz\'alez}\ \emph
  {et~al.}(2012)\citenamefont {Fern\'andez-Gonz\'alez}, \citenamefont {Schuch},
  \citenamefont {Wolf}, \citenamefont {Cirac},\ and\ \citenamefont
  {P\'erez-Garc\'{\i}a}}]{fernandez:1d-uncle}%
  \BibitemOpen
  \bibfield  {author} {\bibinfo {author} {\bibfnamefont {C.}~\bibnamefont
  {Fern\'andez-Gonz\'alez}}, \bibinfo {author} {\bibfnamefont {N.}~\bibnamefont
  {Schuch}}, \bibinfo {author} {\bibfnamefont {M.~M.}\ \bibnamefont {Wolf}},
  \bibinfo {author} {\bibfnamefont {J.~I.}\ \bibnamefont {Cirac}}, \ and\
  \bibinfo {author} {\bibfnamefont {D.}~\bibnamefont {P\'erez-Garc\'{\i}a}},\
  }\href@noop {} {\bibfield  {journal} {\bibinfo  {journal} {Phys.\ Rev.\
  Lett.}\ }\textbf {\bibinfo {volume} {109}},\ \bibinfo {pages} {260401}
  (\bibinfo {year} {2012})},\ \Eprint {http://arxiv.org/abs/arXiv:1111.5817}
  {arXiv:1111.5817} \BibitemShut {NoStop}%
\bibitem [{\citenamefont {Fernandez-Gonzalez}\ \emph
  {et~al.}(2015)\citenamefont {Fernandez-Gonzalez}, \citenamefont {Schuch},
  \citenamefont {Wolf}, \citenamefont {Cirac},\ and\ \citenamefont
  {Perez-Garcia}}]{fernandez-gonzalez:uncle-long}%
  \BibitemOpen
  \bibfield  {author} {\bibinfo {author} {\bibfnamefont {C.}~\bibnamefont
  {Fernandez-Gonzalez}}, \bibinfo {author} {\bibfnamefont {N.}~\bibnamefont
  {Schuch}}, \bibinfo {author} {\bibfnamefont {M.~M.}\ \bibnamefont {Wolf}},
  \bibinfo {author} {\bibfnamefont {J.~I.}\ \bibnamefont {Cirac}}, \ and\
  \bibinfo {author} {\bibfnamefont {D.}~\bibnamefont {Perez-Garcia}},\
  }\href@noop {} {\bibfield  {journal} {\bibinfo  {journal} {Commun.\ Math.\
  Phys.}\ }\textbf {\bibinfo {volume} {333}},\ \bibinfo {pages} {299} (\bibinfo
  {year} {2015})},\ \Eprint {http://arxiv.org/abs/arXiv:1210.6613}
  {arXiv:1210.6613} \BibitemShut {NoStop}%
\bibitem [{\citenamefont {{Bombin}}(2010)}]{bombin:defects}%
  \BibitemOpen
  \bibfield  {author} {\bibinfo {author} {\bibfnamefont {H.}~\bibnamefont
  {{Bombin}}},\ }\href {\doibase 10.1103/PhysRevLett.105.030403} {\bibfield
  {journal} {\bibinfo  {journal} {Phys.\ Rev.\ Lett.}\ }\textbf {\bibinfo
  {volume} {105}},\ \bibinfo {eid} {030403} (\bibinfo {year} {2010})},\ \Eprint
  {http://arxiv.org/abs/1004.1838} {arXiv:1004.1838} \BibitemShut {NoStop}%
\bibitem [{\citenamefont {{Kitaev}}\ and\ \citenamefont
  {{Kong}}(2012)}]{kitaev:gapped-boundaries}%
  \BibitemOpen
  \bibfield  {author} {\bibinfo {author} {\bibfnamefont {A.}~\bibnamefont
  {{Kitaev}}}\ and\ \bibinfo {author} {\bibfnamefont {L.}~\bibnamefont
  {{Kong}}},\ }\href {\doibase 10.1007/s00220-012-1500-5} {\bibfield  {journal}
  {\bibinfo  {journal} {Commun. Math. Phys.}\ }\textbf {\bibinfo {volume}
  {313}},\ \bibinfo {pages} {351} (\bibinfo {year} {2012})},\ \Eprint
  {http://arxiv.org/abs/1104.5047} {arXiv:1104.5047} \BibitemShut {NoStop}%
\bibitem [{\citenamefont {{Barkeshli}}\ \emph {et~al.}(2013)\citenamefont
  {{Barkeshli}}, \citenamefont {{Jian}},\ and\ \citenamefont
  {{Qi}}}]{barkeshli:genons}%
  \BibitemOpen
  \bibfield  {author} {\bibinfo {author} {\bibfnamefont {M.}~\bibnamefont
  {{Barkeshli}}}, \bibinfo {author} {\bibfnamefont {C.-M.}\ \bibnamefont
  {{Jian}}}, \ and\ \bibinfo {author} {\bibfnamefont {X.-L.}\ \bibnamefont
  {{Qi}}},\ }\href {\doibase 10.1103/PhysRevB.87.045130} {\bibfield  {journal}
  {\bibinfo  {journal} {\prb}\ }\textbf {\bibinfo {volume} {87}},\ \bibinfo
  {eid} {045130} (\bibinfo {year} {2013})},\ \Eprint
  {http://arxiv.org/abs/1208.4834} {arXiv:1208.4834} \BibitemShut {NoStop}%
\bibitem [{\citenamefont {{Barkeshli}}\ and\ \citenamefont
  {{Wen}}(2010)}]{barkeshli:u1-times-u1-rtimes-z2}%
  \BibitemOpen
  \bibfield  {author} {\bibinfo {author} {\bibfnamefont {M.}~\bibnamefont
  {{Barkeshli}}}\ and\ \bibinfo {author} {\bibfnamefont {X.-G.}\ \bibnamefont
  {{Wen}}},\ }\href {\doibase 10.1103/PhysRevB.81.045323} {\bibfield  {journal}
  {\bibinfo  {journal} {\prb}\ }\textbf {\bibinfo {volume} {81}},\ \bibinfo
  {eid} {045323} (\bibinfo {year} {2010})},\ \Eprint
  {http://arxiv.org/abs/0909.4882} {arXiv:0909.4882} \BibitemShut {NoStop}%
\bibitem [{Note2()}]{Note2}%
  \BibitemOpen
  \bibinfo {note} {Note that the largest eigenvalue of the transfer operator
  for the normalization {$\delimiter "426830A \psi _{4a}\delimiter "026A30C
  \psi _{4a}\delimiter "526930B $} of the non-abelian anyon $4a$ is degenerate,
  which allows it to have non-zero overlap with more than one sector (while
  without such a degeneracy, i.e., for abelian anyons, the normalized overlap
  must always be zero or one).}\BibitemShut {Stop}%
\bibitem [{Note3()}]{Note3}%
  \BibitemOpen
  \bibinfo {note} {The exact condensation pattern is given by
  $1a+1c+2a+2c+4a+4a \rightarrow \protect \mathaccentV {hat}05E{1}$, with $4a$
  having twice the weight as compared to the other anyons.}\BibitemShut {Stop}%
\end{thebibliography}
\end{document}